\documentstyle[prb,twocolumn,epsfig,amssymb,floats,aps]{revtex}
\begin{document}
\draft

\twocolumn[\hsize\textwidth\columnwidth\hsize\csname @twocolumnfalse\endcsname

\title{Bond-order-wave phase and quantum phase transitions \\ in the 
one-dimensional extended Hubbard model}

\author{Pinaki Sengupta} 
\address{Department of Physics, University of Illinois at 
Urbana-Champaign, 1110 West Green Street, Urbana, Illinois 61801}

\author{Anders W. Sandvik} 
\address{Department of Physics, {\AA}bo Akademi University, Porthansgatan 3,
FIN-20500, Turku, Finland}

\author{David K. Campbell}
\address{Departments of Physics and of Electrical and Computer Engineering, 
Boston University, 44 Cummington Street, \\ Boston, Massachusetts 02215}

\date{\today}

\maketitle

\begin{abstract}
We use a stochastic series expansion quantum Monte Carlo  method to study 
the phase diagram of the one-dimensional extended Hubbard model at half 
filling for small to intermediate values of the on-site ($U$) and 
nearest-neighbor ($V$) repulsions. We confirm the existence of a novel, 
long-range-ordered  bond-order-wave (BOW) phase recently predicted by
Nakamura (J.~Phys.~Soc.~Jpn.~{\bf 68}, 3123 (1999)) in a small region of 
the parameter space between the familiar charge-density-wave (CDW) state for 
$V \agt U/2$ and the state with dominant spin-density-wave (SDW) fluctuations 
for $V \alt U/2$. We discuss the nature of the transitions among these states
and evaluate some of the critical exponents. Further, we determine accurately
the position of the multi-critical point, $(U_m,V_m)= (4.7 \pm 0.1, 2.51 
\pm 0.04)$ (in energy units where the hopping integral is normalized to unity),
above which the two continuous SDW-BOW-CDW transitions are replaced by one 
discontinuous (first-order) direct SDW-CDW transition. We also discuss the
evolution of the CDW and BOW states upon hole doping. We find that in both
cases the ground state is a Luther-Emery liquid, i.e., the spin gap remains 
but the charge gap existing at half-filling is immediately closed upon doping.
The charge and bond-order correlations decay with distance $r$ as 
$r^{-K_\rho}$, where $K_\rho$ is approximately $0.5$ for the parameters we 
have considered. We also discuss advantages of using parallel tempering 
(or exchange Monte Carlo) --- an extended ensemble method that we here 
combine with quantum Monte Carlo  --- in studies of quantum phase 
transitions.
\end{abstract}

\pacs{PACS: 71.10.-w, 71.27.+a, 71.30.+h, 05.30.-d}
\vskip2mm]

\section{Introduction}
\label{sec:intro}

The one-dimensional (1D) extended Hubbard model has been extensively studied 
in recent years, both as an important theoretical test-bed for studying 
novel concepts in 1D (e.g., spin-charge separation) and methods, (e.g., 
quantum Monte Carlo, exact diagonalization, and the density matrix 
renormalization group) and as a useful model for several classes of quasi 
1D materials including copper-oxide materials related to the high-Tc cuprate
superconductors,\cite{cuo} conducting polymers \cite{-ch-ch-} and organic 
charge-transfer salts.\cite{ttftcnq} General 1D extended Hubbard models 
differ from the standard Hubbard model, which includes only an on-site 
electron-electron interaction ($U$), by the addition of longer-range 
interactions which are necessary to explain several experimentally 
observed effects in real materials, e.g., excitons in conducting polymers. 
The simplest extended Hubbard model (henceforth, EHM), on which we focus in 
this article, consists of adding a nearest neighbor interaction $V$.
If the interaction parameters are assumed to arise solely from Coulomb 
interactions, both $U$ and $V$ are repulsive (positive), and $U>V$. However, 
viewed as phenomenological parameters incorporating the effects of additional 
(e.g. electron-phonon) interactions, the ranges of these parameters can 
be much broader, including $U,V<0$. The Hamiltonian is
\begin{eqnarray}
H & = & -t\sum_{i,\sigma}(c^{\dagger}_{i+1,\sigma}c_{i,\sigma} + h.c.) 
\nonumber\\
  & & +U\sum_i(n_{i,\uparrow}-\hbox{$1\over 2$})
(n_{i,\downarrow}-\hbox{$1\over 2$}) 
\nonumber\\
  &   & +V\sum_i(n_{i+1}-1)(n_i-1) - \mu\sum_in_i,
\label{H}
\end{eqnarray}
where $c^{\dagger}_{i,\sigma}(c_{i,\sigma})$ creates (annihilates) an 
electron with spin $\sigma$ at site $i$, $t$ is the hopping integral 
between adjacent sites and $\mu$ is the chemical potential. Henceforth 
we set $t=1$ and express the interaction parameters $U$ and $V$ in 
units of $t$.

The ground state phase diagram of the EHM at half filling $(\mu=0)$ has 
been extensively studied using both analytical and numerical methods. 
Despite the apparent simplicity of the model, the phase diagram shows 
surprisingly rich structure. In the limit $V$=0 (the standard Hubbard 
model), the Hamiltonian (\ref{H}) can be diagonalized exactly using the 
generalized Bethe Ansatz.\cite{liebwu} For $V\neq0$, the model has been 
studied using perturbative methods and numerical simulations.\cite{bari,emery1,solyom,hirsch1,hirsch2,fradkin1,fradkin2,vandongen,fourcade,voit,lin}
Broadly, the phase diagram consists of insulating phases with dominant 
charge-density-wave (CDW) and spin-density-wave (SDW) characters and metallic 
phases where singlet and triplet superconducting correlations dominate. In 
the physically relevant region for "Coulomb-only" parameters ($U,V>0$), the 
system is in a CDW phase for large $V/U$ and in an state with dominant SDW 
fluctuations for small $V/U$. The CDW phase has broken discrete symmetry 
characterized predominantly by alternating doubly occupied and empty sites 
and exhibits long-range order. The SDW phase, on the other hand, has 
continuous symmetry and hence cannot exhibit long-range order in 1D (by 
the Mermin-Wagner theorem). Instead, it is a critical state characterized 
by the slow (algebraic) decay of the staggered spin-spin correlation function.
Indeed, in the limit $U\gg 1$, $U\gg V$, 
the model reduces to an effective Heisenberg 
model with $J\sim 1/(U-V)$. For small $U$ and $V$ $(U,V\ll1)$, the boundary 
between the CDW and the SDW phases was predicted to be at $U=2V$ using weak 
coupling renormalization group techniques (''g-ology'').\cite{emery1,solyom} 
Strong coupling calculations using second-order perturbation theory also 
gave the same phase boundary $(U=2V)$ between the CDW and the SDW phases 
for large $U$ and $V$ $(U,V \gg 1)$.\cite{bari,emery1} 
For intermediate values of the 
parameters, the phase boundary was found to be shifted slightly away from 
the $U=2V$ line such that the SDW phase is enhanced, as shown by quantum
Monte Carlo simulations \cite{hirsch1,hirsch2} as well as strong coupling 
calculations using perturbation theory up to the fourth order.\cite{vandongen} 
Moreover, the nature of the transition is quite different in the two coupling 
regions, changing from continuous (second-order) in the weak coupling limit to 
discontinuous (first-order) in the strong coupling limit. Estimates for the 
location of the multi-critical point, where the nature of the transition 
changes, have ranged from $U_m \simeq 1.5$ to $U_m \simeq 5$ (and $V_m \approx
U_m/2$).\cite{hirsch1,hirsch2,fradkin1,fradkin2,voit} Despite the broad 
uncertainty in the actual value of the tricritical point, the phase diagram 
was believed to be well understood.

\begin{figure}
\centering
\epsfxsize=8.3cm
\leavevmode
\epsffile{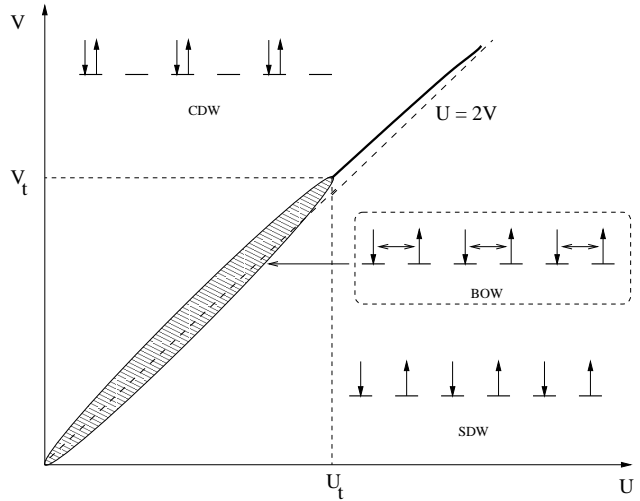}
\vskip1mm
\caption{Schematic ground state phase diagram of the EHM at half filling, 
as proposed by Nakamura. The CDW and BOW phases are long-range-ordered 
(broken-symmetry), whereas the SDW phase has no broken symmetry but exhibits 
an algebraically decaying spin-spin correlation function.} 
\label{gs}
\end{figure}

Recently, however, by studying the EHM ground state broken symmetries using 
level crossings in excitation spectra obtained by exact diagonalization, 
Nakamura \cite{nakamura} has argued for the existence of a novel 
bond-order-wave (BOW) phase for small to intermediate values of $U$ and 
$V$ in a narrow strip between the CDW and the SDW phases. The BOW phase is 
characterized by alternating strengths of the expectation value of the 
kinetic energy operator on the bonds. 
It is predicted to be a state where the discrete (two-fold) symmetry is broken
and should hence exhibit true long-range order. Nakamura thus argues that the 
transition between CDW and SDW phases in this region is replaced two 
separate transitions: $(i)$ a continuous transition from CDW to BOW; and 
$(ii)$ a Kosterlitz-Thouless spin-gap transition from BOW to SDW. The BOW 
region vanishes at the multi-critical point beyond which the transition 
between CDW and SDW phases is direct and discontinuous. A schematic phase 
diagram including Nakamura's BOW state is shown in Fig.~\ref{gs}.

Considering the long history of the 1D EHM and the large number of studies 
of the $U \approx 2V$ region with a variety of analytical and numerical tools, 
the proposal of a new phase is certainly remarkable. Importantly, the level 
crossing method used by Nakamura cannot by itself exclude the conventional 
scenario of a direct SDW-CDW transition for the whole range of $U,V > 0$; a 
level crossing corresponding to this transition was also found \cite{nakamura} 
between the SDW-BOW and BOW-CDW crossing curves. The position of the BOW-CDW 
level crossing is, however, in closer agreement with the strong-coupling 
result for the vanishing of the CDW order, and this was taken as evidence of
a long-range-ordered BOW in the ground state for certain parameters. It is 
important to confirm this hitherto undiscovered phase using other methods. 

To attempt this confirmation, we have used the highly efficient
stochastic series expansion (SSE) quantum Monte Carlo method 
\cite{anders1,anders2,sseloop} to study the EHM at half filling in the 
vicinity of $U=2V$. This method allows us to probe directly the spin- charge- 
and bond-order correlations in the ground state of lattices with more than 
one hundred sites (up to 256 sites were used in this study). Using finite-size 
scaling techniques for the various order parameters, we confirm the existence 
of a BOW state with spin and charge gaps in a region very close to that 
predicted by Nakamura for small $U,V$. We also further improved the SSE 
simulations by applying a quantum version of the thermal parallel tempering 
scheme (or exchange Monte Carlo) \cite{tempering1,tempering2,tempering3}
for simulations close to and across the phase boundaries. This ``quantum 
parallel tempering'' greatly reduced the effects of ``sticking'' --- where 
the simulation gets trapped in the wrong phase close to a phase boundary --- 
and was found to be particularly useful for the discontinuous (first-order) 
direct SDW-CDW transition. As a consequence, we were able to obtain a more
accurate estimate for the location of the multi-critical point $(U_m,V_m)$ 
where the BOW phase vanishes and is replaced by a first-order SDW-CDW 
transition line. As we discuss below, we find $U_m=4.7 \pm 0.1, V_m=2.51 
\pm 0.04$.

In order to investigate the possibility of soliton lattices forming out of
the long-range CDW and BOW states when doping away from half-filling, we
have also carried out some simulations of lightly doped systems. We find
that the in both cases the ground state is a Luther-Emery liquid, with
a spin gap and slow algebraic decay ($\sim r^{-K_\rho}$, with $K_\rho
\approx 0.5$) of the dominant CDW and BOW correlations.

The remainder of the paper is organized into four sections and two
appendices. In Sec.~\ref{sec:methods} we briefly sketch the SSE method and 
introduce the different observables we study. In Sec.~\ref{sec:results} we 
present the results of our simulations at half-filling and discuss their 
interpretation. Doped systems are considered in Sec.~IV. We conclude with a 
brief summary in Sec.~\ref{sec:summary}. In Appendix A we present some 
important details of the extension of the SSE method to allow efficient 
loop updates for fermions. We illustrate the advantages of the quantum 
parallel tempering scheme in Appendix B.

\section{Numerical Methods and Observables}
\label{sec:methods}

\subsection{The SSE method and its fermion loop-update extension}

The SSE method \cite{anders1,anders2}
is a finite-temperature quantum Monte Carlo method based on importance
sampling of the diagonal elements of the Taylor expansion of $e^{-\beta H}$, 
where $\beta$ is the inverse temperature $\beta=t/T$. Ground state expectation
values can be obtained using sufficiently large values of $\beta$, and there
are no approximations beyond statistical errors. Recently, in the context
of spin systems,\cite{sseloop} an efficient ``operator loop update'' was 
developed to sample the operator sequences appearing in the expansion. The 
resulting method has proven to be very efficient for several different
models.\cite{bilayer,wessel,dorneich} To apply the most efficient variant 
of SSE method to the EHM, we need to generalize the previous operator loop 
update scheme to spinful fermions. This is an important extension, but because
of its technical nature we have relegated our detailed discussion of it 
to an appendix.

We have applied the SSE algorithm to the 1D EHM for system sizes ranging from 
$N = 8$ to $256$ sites, with maximum inverse temperatures $\beta$ chosen 
appropriately to isolate the ground state. We have verified the correctness
of the simulation code by comparing $N=8$ results with exact diagonalization 
(Lanczos) results.  

Although the operator-loop update is indeed significantly more efficient than
previous local updates for sampling of the SSE configurations, we still
have problems with ``trapping'' close to a first-order phase transition,
i.e., the simulation can get stuck in the wrong phase very close to the 
critical point. There are also problems with slow dynamics in long-range 
ordered phases with a broken discrete symmetry (such as BOW or CDW phases).
In order to overcome these problems we have developed a ``quantum parallel 
tempering'' scheme --- a generalization of the thermal parallel tempering 
method \cite{tempering1,tempering2,tempering3} commonly used to equilibrate 
classical spin glass simulations. The method amounts to running several 
simulations on a parallel computer, using a fixed value of $U$ and different
closely spaced values of $V$ at and around the critical value $V_c$. Along 
with the usual Monte Carlo updates, we attempt to swap the configurations 
for processes with adjacent values of $V$ at regular intervals (typically 
after every Monte Carlo step) according to a scheme that maintains
detailed balance in the space of the parallel simulations, as explained
in Appendix B. In contrast to Ref.~\onlinecite{tempering3}, we here find 
parallel tempering to be particularly useful in the study of the first-order 
transition, where the problem of trapping is the most pronounced. 
In Appendix B we also present a comparative example to illustrate the 
improvement obtained by parallel tempering. 

\subsection{Observables}

In addition to the ground state energy, $E=\langle H\rangle/N$, the 
observables we study include the static structure factors and susceptibilities
corresponding to the different phases (CDW, SDW and BOW). The structure 
factors are given by
\begin{eqnarray}
S_{SDW}(q) &=& \frac{1}{N} \sum_{j,k}e^{iq(j-k)}\langle
S_j^zS_k^z\rangle,\nonumber\\
S_{CDW}(q) &=& \frac{1}{N} \sum_{j,k}e^{iq(j-k)}
           \langle n_jn_k\rangle - \langle n_j\rangle^2,\nonumber\\
S_{BOW}(q) &=& \frac{1}{N} \sum_{j,k}e^{iq(j-k)}
\langle k_jk_k\rangle - \langle k_j\rangle^2,  
\label{eq:sf}
\end{eqnarray}
\noindent where 
\begin{equation}
k_j=\sum_{\sigma=\uparrow,\downarrow}(c^{\dagger}_{j+1,\sigma}c_{j,\sigma} 
+ h.c.)
\end{equation}
is the kinetic energy operator associated with the $j$th bond. 
The corresponding static susceptibilities are given by
\begin{equation}
\chi_{SDW}(q) = \frac{1}{N} 
\sum_{j,k}e^{iq(j-k)}\int^\beta_0 d\tau \langle S^z_j(\tau) S^z_k(0)\rangle 
\label{eq:xcdw}
\end{equation} 
\noindent and analogous expressions for $\chi_{CDW}(q)$ and 
$\chi_{BOW}(q)$. Since all the phases mentioned have a period 2, the 
staggered structure factor and susceptibilities are the most important 
observables.  We define order parameters for the phases in terms of the 
staggered structure factors:
\begin{equation}
m_{\alpha}=\sqrt{S_\alpha(\pi)/N}, 
\label{eq:m2}
\end{equation}
\noindent where $\alpha = $ CDW, SDW, or BOW. We have 
also studied the charge stiffness constant, $\rho_c$. It is defined as the
second derivative of the internal energy per site, $E$, with respect to a 
twist, $\phi$,\cite{kohn}
\begin{equation}
\rho_c = \frac{\partial ^2E(\phi)}{\partial\phi ^2}, 
\label{eq:rhoc}
\end{equation} 
under which the hopping term in the Hamiltonian (\ref{H}) is 
replaced by 
\begin{equation}
k_c(\phi) = -t\sum_{j,\sigma}(e^{-i\phi}c^{\dagger}_{j+1,\sigma}c_{j,\sigma}
+ h.c.).
\end{equation} 
The spin stiffness constant, $\rho_s$, is defined by a similar expression, 
with the hopping term now being replaced by 
\begin{equation}
k_s(\phi) = -t\sum_{j,\sigma}(e^{-i\phi_{\sigma}}
c^{\dagger}_{j+1,\sigma}c_{j,\sigma} + h.c.), 
\end{equation} 
with $\phi_{\uparrow}=-\phi_{\downarrow}=\phi$. In the framework of the 
SSE method, the estimators for the charge and spin stiffness are given 
in terms of expectation values of squared winding numbers (see Appendix A).

\section{Results at half-filling}
\label{sec:results}

As noted above, we have studied chains with $N$ up to $256$ with periodic 
boundary conditions.\cite{periodicnote} Typically, an inverse temperature 
of $\beta=2N$ was sufficient for the calculated properties to have converged 
to their ground state values, except in the case of $N=256$, for which 
$\beta=4N$ was needed for some quantities. In this section we first discuss 
our evidence for the existence of a long-range BOW phase, then our analysis
of the continuous BOW-CDW and SDW-BOW transitions for small $(U,V)$, the
discontinuous SDW-CDW transition for large $(U,V)$, and finally our
determination of the location of the multi-critical point separating 
these transitions. 

\subsection{Existence of the BOW phase}

\begin{figure}
\centering
\epsfxsize=8cm
\leavevmode
\epsffile{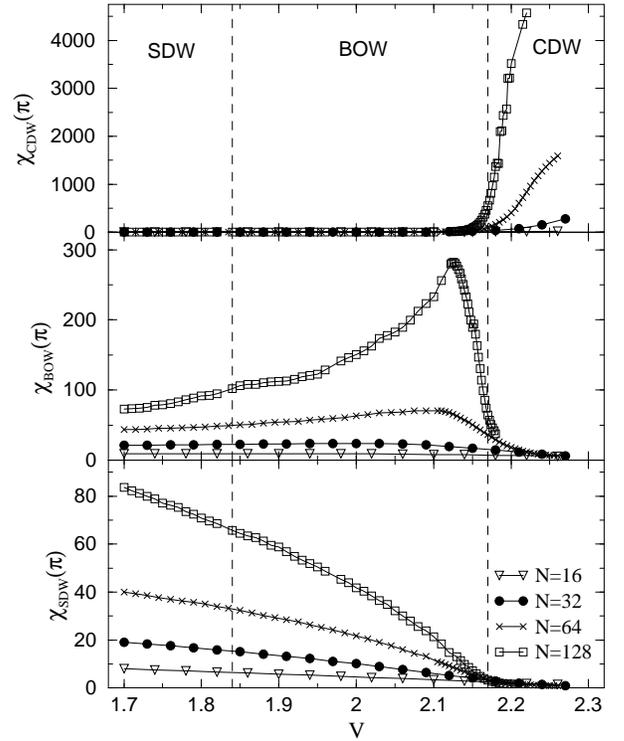}
\vskip1mm
\caption{The variation with $V$ (at fixed $U=4$) of the staggered 
susceptibilities (CDW, BOW, and SDW, from the top) in the neighborhood
of the BOW phase predicted by Nakamura (the vertical dashed lines show the 
predicted SDW-BOW and BOW-CDW boundaries). The statistical errors are 
typically of the order of the size of the symbols 
(slightly larger for the $N=128$ CDW 
at high $V$). The scans for $N=16$ and $32$ were obtained in single parallel 
tempering simulations, whereas those for $N=64$ and $128$ consisted of two
and four non-overlapping runs, respectively.} 
\label{bowscan}
\end{figure}

Plots of the variation of the staggered susceptibilities corresponding to the 
three different phases --- CDW, SDW, and BOW ---  show the existence of strong 
BOW fluctuations in a region with $V \simeq U/2$ in parameter space where 
Nakamura predicted a BOW state. Fig.~\ref{bowscan} 
is such a plot for $U=4$ and $1.7 \le V < 2.3$. In a
long-range ordered phase (BOW, CDW), the corresponding $\chi(\pi)$ is expected
to diverge with increasing system whereas the other two susceptibilities 
should converge to constants. In the SDW phase there is no long-range order
but algebraically decaying correlations of both SDW and BOW nature; hence
$\chi_{SDW}(\pi)$ and $\chi_{BOW}(\pi)$ should both diverge here, but the
BOW divergence should be much slower than in the long-ranged BOW phase. These 
behaviors are indeed seen in Fig.~\ref{bowscan}, with the susceptibilities for 
SDW, BOW, and CDW dominating in turn as $V$ is increased. The BOW-CDW phase 
boundary can be quite well resolved since it involves a standard second order 
(continuous) phase transition. On the other hand, the SDW-BOW boundary
is more difficult to locate, for it involves a Kosterlitz-Thouless transition
in which the spin gap opens exponentially slowly as one enters the BOW phase,
\cite{nakamura} resulting in only a slow decay of the staggered SDW 
susceptibility in the BOW phase for the system sizes accessible in 
our work.

\begin{figure}
\centering
\epsfxsize=8.3cm
\leavevmode
\epsffile{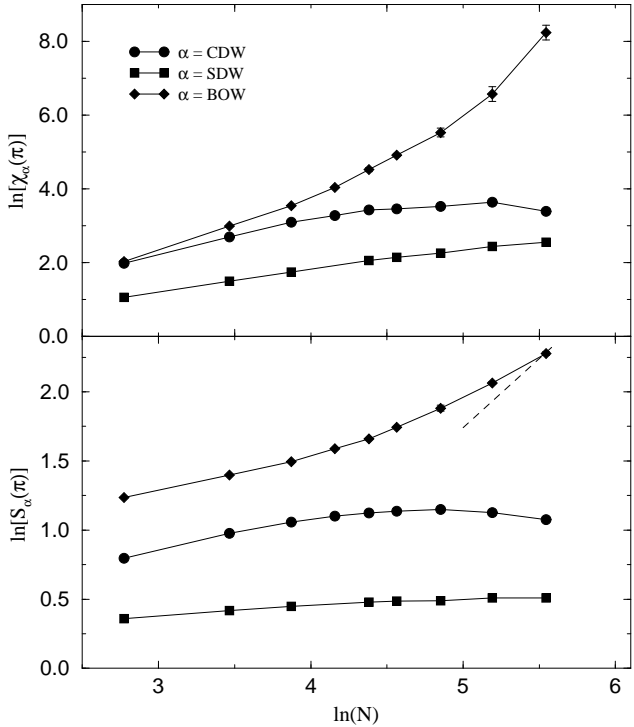}
\vskip1mm
\caption{ln$[\chi(\pi)]$ and ln$[S(\pi)]$ vs ln$[N]$ for the different 
phases at $U=4$, $V=2.14$ and system sizes $N$ up to $256$. The dashed
line in the $S(\pi)$ panel has slope $1$.} 
\label{bowproof}
\end{figure}

Fig.~\ref{bowproof} shows ln$[\chi_\alpha (\pi)]$ 
and ln$[S_\alpha (\pi)]$ versus ln$[N]$ 
for the parameters $(U,V)=(4,2.14)$ for which the ground state should be 
inside the BOW phase. We find that both $\chi_{BOW}(\pi)$ and $S_{BOW}(\pi)$ 
diverge strongly with system size, whereas the structure factor and 
susceptibility corresponding to CDW have a maximum and then decrease with 
system size for large $N$. The SDW structure factor appears to have converged 
for $N=256$ but the susceptibility still shows a weak growth --- in a 
spin-gapped BOW phase it should eventually converge, too, but if the gap is 
very small the convergence occurs only for much larger systems. The growth
with $N$ seen here is much slower than $~N$, which should be the asymptotic 
behavior in an SDW phase for any spin-rotationally invariant 1D system,
\cite{ll} and the growth slows with increasing $N$. Hence an asymptotic 
divergence of $\chi_{SDW}(\pi)$ can be excluded. The dominant asymptotic 
characteristic of the ground state is clearly BOW. The system sizes considered
are not large enough for $S_{BOW}(\pi)$ to have reached the asymptotic
behavior $\sim N$ expected if there is long-range order, which we will
explain further below. The very fast divergence of $\chi_{BOW}(\pi)$ is 
expected on account of the two-fold degenerate BOW ground state. For finite 
$N$ this degeneracy is not perfect, but an exponentially fast closing of the 
gap between the symmetric and antisymmetric linear combinations of the two 
asymptotically degenerate symmetry-broken ordered states can be expected, 
which would eventually cause $\chi_{BOW}(\pi)$ to diverge exponentially. 

\begin{figure}
\centering
\epsfxsize=8.3cm
\leavevmode
\epsffile{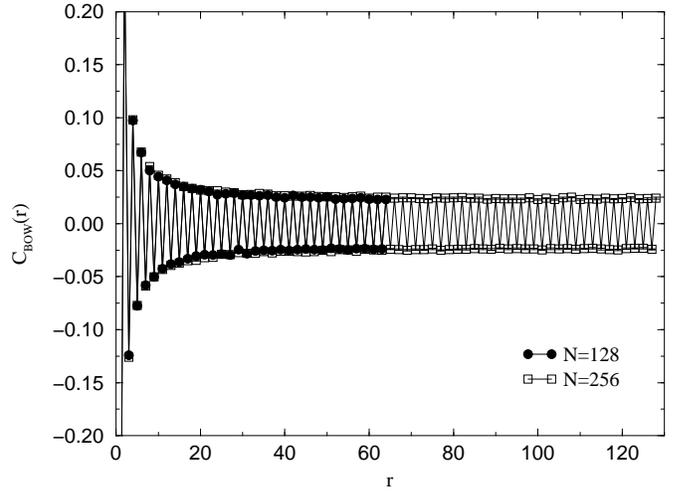}
\vskip1mm
\caption{The real-space BOW correlation function at $U=4$, $V=2.14$
for system sizes $N=128$ and $256$.} 
\label{bowr}
\end{figure}

The most direct evidence for a long-range BOW comes from the the real-space
kinetic energy correlation function
\begin{equation}
C_{BOW}(r) = {1\over N}\sum\limits_{i=1}^N
\langle k_ik_{i+r}\rangle - \langle k_i\rangle^2.
\end{equation}
As seen in Figure \ref{bowr}, this correlation function oscillates with 
period 2 and its magnitude decays considerably for short distances. For long 
distances there is a convergence to a constant, non-zero magnitude, which
is the same within statistical errors for $N=128$ and $256$. The significant
enhancement of the correlations at short distances explain the deviations
from the expected asymptotic linear scaling of the integrated correlation 
function, $S_{BOW}(\pi)$, for the system sizes shown in 
Figure \ref{bowproof}.

Further proof of the existence of the BOW phase is obtained by looking 
for spin and charge gaps in this region. Instead of calculating the gaps
directly, which can not easily be done to high accuracy for large system 
sizes, we use the following indirect method: It is known~\cite{ll} that if 
the ground state of a 1D system is gapless in the spin sector, the Luttinger
liquid parameter $K_\sigma$ governing the asymptotic equal-time spin 
correlation function is $K_{\sigma}=1$.\cite{voit} It has been further 
shown~\cite{torsten} that the slope $S_{SDW}(q)/q$ gives $K_{\sigma}/\pi$ in 
the limit $q\rightarrow 0$. Hence, $S_{SDW}(q)/q\rightarrow 1/\pi$ as 
$q\rightarrow 0$. On the other hand, if the ground state has a spin gap, 
$S_{SDW}(q)/q \rightarrow 0$ as $q\rightarrow 0$. With this criterion even 
a very small spin gap can be detected, since it is in practice sufficient to 
see that $\pi S_{SDW}(q)/q$ decays below $1$ for small $q$ to conclude
that $K_\sigma \not =1$ and hence that a spin gap must be present. Similarly, 
for a ground state with no charge gap, $\pi S_{CDW}(q)/q\rightarrow 
K_{\rho}$ as $q\rightarrow 0$ whereas if the ground state does have a 
charge gap, $S_{CDW}(q)/q \rightarrow 0$ as $q\rightarrow 0$. Unlike 
$K_{\sigma}$, where the value is fixed at $1$ for spin rotationally invariant 
systems, the Luttinger liquid charge correlation parameter $K_\rho$ is a 
function of $U$ and $V$, and its precise value for given $U$ and $V$ is 
not known (except at $V=0$ \cite{schulz}). Due to the logarithmic corrections 
typical for 1D systems, it is very difficult to observe numerically that 
$\pi S_{SDW}(q)/q$ becomes exactly $1$.\cite{eckle,eggert1,eggert2} 
Empirically, we have found that in the gapless case the value $1$ is always 
approached from above (which is the case also for spin systems \cite{eggert1}),
and hence the detection of the spin gap using this quantity is not hampered 
by the log-corrections --- if $\pi S_{SDW}(q)/q$ decays below $1$ one can 
conclude that here is a gap. 

\begin{figure}
\centering
\epsfxsize=8cm
\leavevmode
\epsffile{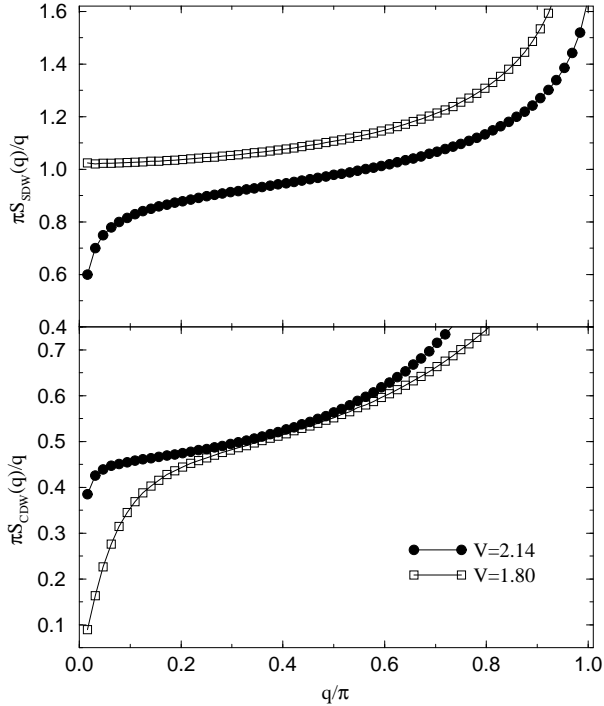}
\vskip1mm
\caption{$S_{SDW}(q)/q$ and $S_{CDW}(q)/q$ vs $q$ for $U=4$ and $V=2.14$ 
and $V=1.8$ ($N=128$).} 
\label{sq}
\end{figure}

Fig.~\ref{sq} shows $\pi S_{SDW}(q)/q$ and 
$\pi S_{CDW}(q)/q$ versus $q/\pi$ for $U=4$ and two values of $V$. One 
of the points ($V=2.14$) is inside the BOW phase, whereas the other ($V=1.8$) 
is in the SDW phase. The $\pi S_{SDW}(q)/q$ curve for $V=1.8$ is close to 
$1$ for a wide range of $q$ values, whereas the $V=2.14$ curve exhibits a 
sharp drop as $q\rightarrow 0$ indicating, respectively, the absence and the 
presence of a spin gap. Similarly, the evidence for a vanishing limit of 
$S_{CDW}(q)/q$ and hence of a charge gap for $V=1.8$ is clear. Since the 
point $V=2.14$ is quite close to the critical point ($V_c=2.16$) where the 
charge gap vanishes, the magnitude of the gap is very small and we need to 
go to still smaller $q$, i.e., larger system size, to see a pronounced effect
like that for $V=1.8$. Nevertheless, the downturn for the smallest $q$ is 
a good indication of a gap. 

\begin{figure}
\centering
\epsfxsize=8cm
\leavevmode
\epsffile{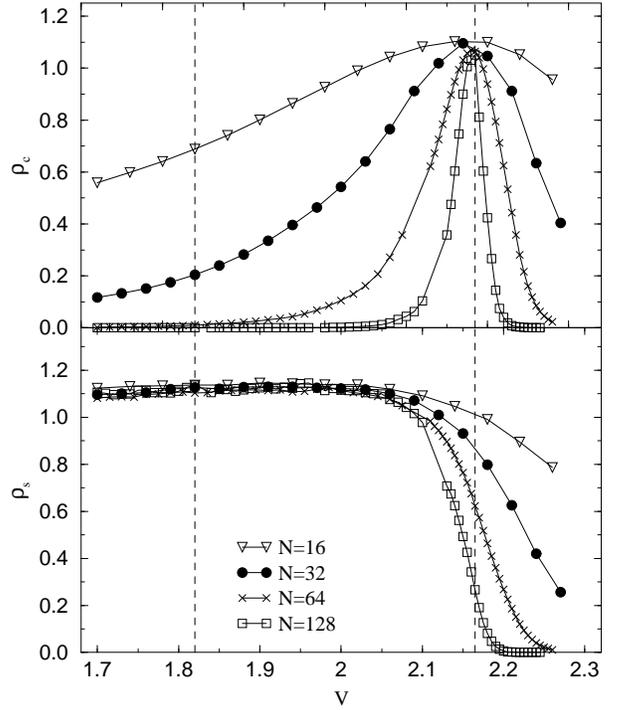}
\vskip1mm
\caption{Behavior of the charge and spin stiffness across the BOW-CDW boundary
for $U=4$. The upper(lower) panel shows the charge(spin) stiffness. The 
vertical dashed lines indicate the position of the phase boundaries according 
to Nakamura.} 
\label{rhocs}
\end{figure}

The opening of spin and charge gaps can also be detected in the spin and
charge stiffness constants, which should vanish as $N \to \infty$ if there
are gaps. The asymptotic charge stiffness should hence be non-zero only 
exactly at the BOW-CDW phase boundary. The spin stiffness should be non-zero 
in the SDW phase, should approach a constant value exactly at the phase
boundary (with logarithmic size-corrections),\cite{ktnote} and vanish inside 
the CDW phase. In Fig.~\ref{rhocs} we show the stiffness constants for 
$U=4$ in the neighborhood of the BOW phase. As expected, the charge 
stiffness peaks at the BOW-CDW phase boundary and decreases rapidly away 
from it, confirming the vanishing of the charge gap only at the phase 
boundary. The peak becomes very sharp for large system
sizes, and the finite-size corrections to its location are small. We find 
this the most accurate way to locate the BOW-CDW phase boundary.
The spin stiffness is clearly zero in the CDW phase, 
and a sharp decrease with increasing $N$ is also seen for $V$ values well 
inside the BOW phase. Since the spin gap opens up exponentially slowly at 
the SDW-BOW boundary it is difficult to locate the transition this way. Our 
data nevertheless indicate that the BOW phase at $U=4$ may not extend down 
to the value $V\approx 1.82$ obtained by Nakamura. We will discuss this
phase transition and determine the transition point more accurately below,
in Sec.~III-C.

\subsection{BOW-CDW transition}

In addition to proving the existence of the BOW phase, we have studied in 
detail the nature of the continuous BOW-CDW transition for 
two different values of $U$ ($U<U_m$). For $(U,V)=(U_c,V_c)$, i.e., 
on the BOW-CDW phase boundary, the real space staggered charge and 
kinetic energy correlation functions fall off algebraically as
\begin{eqnarray}
\langle n_in_{i+r}\rangle (-1)^r &\sim& r^{-\eta},\nonumber \\
(\langle K_iK_{i+r}\rangle - \langle K_i\rangle^2)(-1)^r &\sim& r^{-\eta}.
\label{eq:eta}
\end{eqnarray}
Based on conformal field theory calculations for similar phase transitions 
in 1D spin systems,\cite{nomura} the exponent $\eta$ can be expected to depend 
on $(U_c,V_c)$ but should be the same for both the CDW and BOW correlations. 
This gives the finite-size scaling of the structure factor and the 
susceptibility at the critical point:
\begin{eqnarray}
S_{CDW,BOW}(\pi) &\sim & N^{1-\eta},\nonumber \\ 
\chi_{CDW,BOW}(\pi) &\sim & N^{2-\eta}. 
\label{eq:fss}
\end{eqnarray}
With a spin gap but no charge gap, as was demonstrated above, we expect the 
critical state to be of the Luther-Emery liquid type.\cite{le} The exponent 
$\eta$ is then related to the Luttinger liquid parameter $K_\rho$ by 
$\eta=1-K_\rho$.

\begin{figure}
\centering
\epsfxsize=8.3cm
\leavevmode
\epsffile{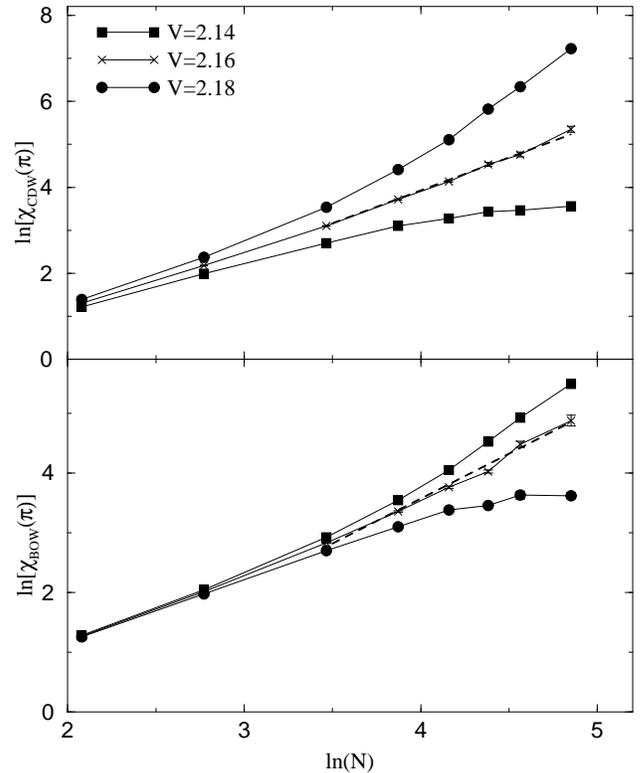}
\vskip1mm
\caption{ln$[\chi_{CDW}(\pi)]$ and ln$[\chi_{BOW}(\pi)]$ vs ln$[N]$ for 
$U$=4 and different values of $V$ near the critical point. The dashed 
lines are fits to the $V=2.16$ data.} 
\label{u4x}
\end{figure}

\begin{figure}
\centering
\epsfxsize=8.3cm
\leavevmode
\epsffile{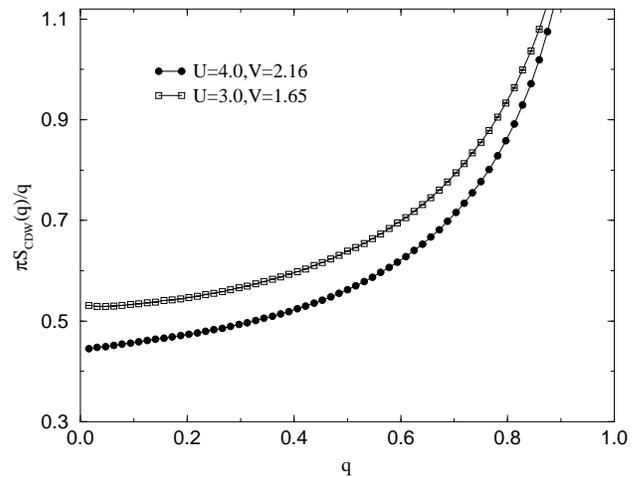}
\vskip1mm
\caption{$\pi S_{CDW}(q)/q$ versus $q/\pi$ 
for two points on the BOW-CDW boundary.} 
\label{krho}
\end{figure}

Fig.~\ref{u4x} presents plots of ln$[\chi_{CDW}]$ and ln$[\chi_{BOW}]$ versus 
ln$[N]$ for $U=4$ and three different values of $V$ around the critical 
point, which as discussed above should be close to $2.16$. The data points for
$V$=2.16 indeed fall almost on straight lines, indicating critical scaling for
both the CDW and BOW fluctuations. The value of the critical exponent $\eta$, 
obtained from the slope of the $V=2.16$ curves for both $\chi_{CDW}$ and 
$\chi_{BOW}$ is $\eta \approx 0.5$. The scaling of the structure factors, 
$S_{CDW}$ and $S_{BOW}$, at $V=2.16$ is also consistent with $\eta \approx 
0.5$. It is, however, difficult to extract a precise value for $\eta$ from
this finite-size scaling, due to subleading corrections to the
scaling, as well as effects from the fact that the $U,V$ point studied is
not exactly on the phase boundary. As was discussed in Sec.~III-A, the 
Luttinger liquid parameter $K_\rho$ can also be extracted from the $q\to 0$ 
limit of $S_{CDW}(q)/q$. This is in general a more accurate method, since 
the convergence with system size is faster for the subleading $1/r^2$ 
contribution to the correlation function which this estimator accesses.
\cite{torsten,schulz} Fig.~\ref{krho} shows results for
$U=4$ and $U=3$ and the respective critical $V$-values. The $q\to 0$ behavior 
gives $K_\rho=0.44 \pm 0.01$ for $U=4$, i.e., $\eta=0.56 \pm 0.01$, which 
hence is consistent with the finite-size scaling of the $q=\pi$ quantities. 
For $U=3$ we obtain $V_c = 1.65$, in agreement with Nakamura's result,
\cite{nakamura} and the critical exponent $\eta=0.47 \pm 0.01$. 

\subsection{SDW-BOW transition}

\begin{figure}
\centering
\epsfxsize=8.3cm
\leavevmode
\epsffile{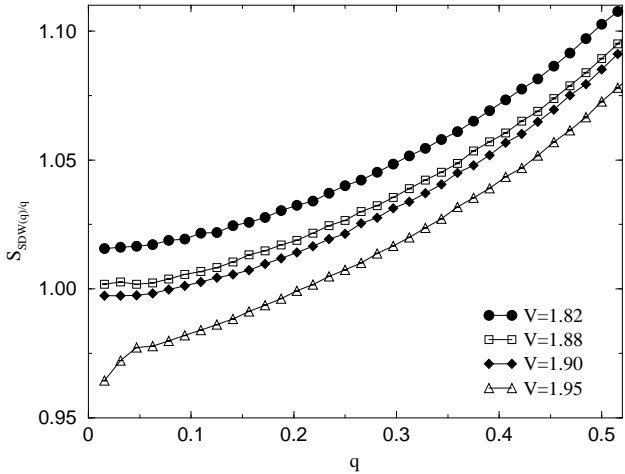}
\vskip1mm
\caption{$S_{SDW}(q)/q$ vs. $q$ for $U=4$ and 4 different values of $V$
around the SDW-BOW boundary.} 
\label{sdwbowsq}
\end{figure}

The SDW-BOW transition is marked by the opening of a spin-gap in the
electronic energy spectrum. As argued by Nakamura, it is a quantum phase
transition of the Kosterlitz-Thouless type and therefore the gap opens up 
exponentially slowly. This makes it difficult to determine the phase 
boundary numerically. The numerical data is affected by large finite-size 
effects that persist up to very large system sizes. As discussed in 
Sec.~\ref{sec:results}A, the most reliable evidence of the existence of a 
spin-gap is obtained from the behavior of $S_{SDW}(q)/q$ as $q\rightarrow 0$.
In practice, an asymptotic value of $\pi S_{SDW}(q)/q < 1$ as $q\rightarrow 0$ 
in any (large) system is an indication of the presence of a spin-gap in the 
thermodynamic limit. This allows us to detect the presence of very small spin 
gaps. Fig.~\ref{sdwbowsq} shows the behavior of $\pi S_{SDW}(q)/q$  
for $U=4$ and different values of $V$. In the gapless
region, logarithmic corrections \cite{eggert1} makes it difficult to observe 
the approach to $1$ as $q\to 0$. In analogy with spin systems,\cite{eggert2} 
we expect the leading log corrections to vanish at the point where the spin 
gap opens, and therefore exactly at the critical point there should be a 
clear scaling to $1$. An apparent reduction of the log correction is indeed
seen in Fig.~\ref{sdwbowsq} as $V$ is increased towards $\approx 1.88$. Based 
on the results, we estimate the SDW-BOW boundary to be at $V=1.89\pm 0.01$ at 
$U=4$. This is slightly higher than Nakamura's critical value $V=1.82$ for 
this $U$. We believe the difference is due to non-asymptotic finite-size 
effects in the exact diagonalization calculation, which used system sizes 
only up to $N=14$. Hence, we find that the BOW phase exists in a slightly 
smaller, while still significant, region of the phase space.

\subsection{First-order SDW-CDW transition}

\begin{figure}
\centering
\epsfxsize=8.3cm
\leavevmode
\epsffile{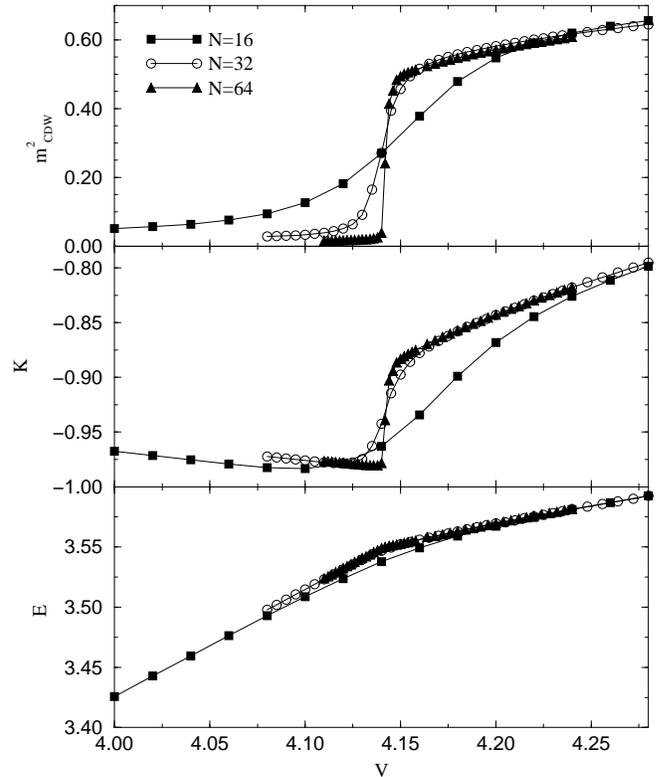}
\vskip1mm
\caption{The behavior of the CDW order parameter, the kinetic energy and 
the ground state energy across the SDW-CDW transition for various system 
sizes and $U=8$.} 
\label{u8}
\end{figure}

For $U>U_m$, the transition is a discontinuous (first-order) direct 
SDW-CDW transition with no intervening BOW phase. Fig.~\ref{u8} shows the 
$V$ dependence of the CDW order parameter, the total energy, and the kinetic 
energy across the phase boundary for $U=8$, which according to previous
studies \cite{hirsch1,hirsch2,fradkin1,fradkin2,voit} should be well within 
the regime of first-order transitions. The characteristics of a first-order 
transition are indeed quite apparent. The order parameter and the kinetic 
energy change rapidly at the transition point, $V_c \approx 4.14$. The 
finite-size effects diminish with increasing $N$ as the results 
approach the limiting behavior of a discontinuity in the order parameter 
and the kinetic energy in the thermodynamic limit. The total energy remains 
continuous, but there is a clear break in slope at the transition.  

\begin{figure}
\centering
\epsfxsize=8.3cm
\leavevmode
\epsffile{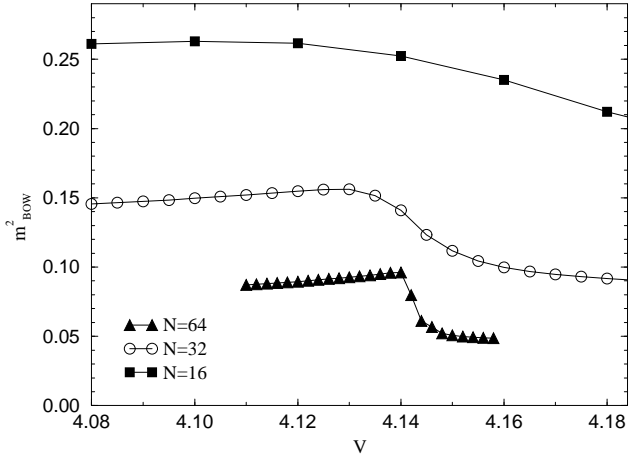}
\vskip1mm
\caption{The behavior of the BOW order parameter across the SDW-CDW transition 
for various system sizes and $U=8$.} 
\label{fig:mbow}
\end{figure}

The size dependence of the BOW order-parameter is shown in Fig.~\ref{fig:mbow}.
It becomes considerably smaller inside the CDW phase than before the 
transition. This is expected, since in the SDW phase, but not in the CDW 
phase, there should be power-law decaying BOW correlations. The BOW order 
parameter decays rapidly with the system size, however, confirming that there 
is no long-range BOW for this $U > U_m$.

The behavior with increasingly sharp discontinuities seen in Fig.~\ref{u8}
and \ref{fig:mbow} 
indicates a first-order transition due to an avoided level crossing. Note
that with increasing chain length the CDW order parameter approaches its 
thermodynamic value from above for $V<V_c$ and from below for $V$ just 
above $V_c$. The curves for different system sizes cross one another in the 
neighborhood of $V=V_c$ and then once again for a higher $V$. The second 
crossing point moves down towards the first one as $N$ increases, whereas 
the first crossing does not change much with $V$ and appears to be a 
good criterion for locating the transition point. 

The two curve crossings can be understood as follows: In a transition caused 
by an avoided level crossing, a crossing of the order parameter curves close 
to the critical coupling (approaching the critical coupling as $N \to \infty$) 
can be expected since the low-energy levels corresponding to an ordered and 
disordered state swap characters within a a parameter range $V \pm \Delta_V 
(N)$, with $\Delta_V(N) \to 0$ as $N \to \infty$. This behavior is seen 
clearly in Fig.~\ref{u8}. The finite-$N$ ground state starts to develop CDW 
characteristics at $V - \Delta_V (N)$ and thus, for a fixed $V < V_c$, the CDW
order parameter decreases with increasing $N$. An analogous argument for fixed
$V > V_c$ close to $V_c$ gives that the in this case the CDW order must 
increase with increasing $N$. On the other hand, for $V \gg V_c$ the 
real-space CDW correlations are enhanced at short distances (in the same way 
as the BOW correlations shown in Fig.~\ref{bowr}) and for small system sizes 
there is also some enhancement of the long-distance correlations due to the 
periodic boundary conditions.\cite{pericorr} Hence, one can expect the CDW 
order parameter, when defined and measured in terms of its squared expectation
Eq.~(\ref{eq:m2}), to again {\it decrease} with $N$ for $V \gg V_c$ and this 
explains the second crossing of the order parameter curves seen in 
Fig.~\ref{u8}.

\subsection{Multi-critical point} 

Although the existence of the tricritical point (which, in vew of the 
existence of the BOW phase, we refer to as the multi-critical 
point) separating the first-order and continuous transition to the CDW
state has long been known, its location in the $(U,V)$ plane has not 
previously been determined accurately using large system sizes. 
Hirsch~\cite{hirsch1,hirsch2} estimated a value of $U_m=3$ using world 
line Monte Carlo. Cannon and Fradkin\cite{fradkin1} obtained $U_m=1.5$ using 
field theory techniques and world lines. Later Cannon, Scalettar and 
Fradkin\cite{fradkin2} obtained a value of $U_m=3.5-5$ using finite-size 
scaling of Lanczos results. Using a combination of bosonization and RG 
techniques, Voit~\cite{voit} obtained $U_m=4.76$. However, as Voit also 
pointed out, the validity of bosonization and RG, which are applicable in 
the limit $U,V \rightarrow 0$, for intermediate values of the parameters 
is {\it a priori} questionable.

\begin{figure}
\centering
\hskip -2mm
\epsfxsize=8.45cm
\leavevmode
\epsffile{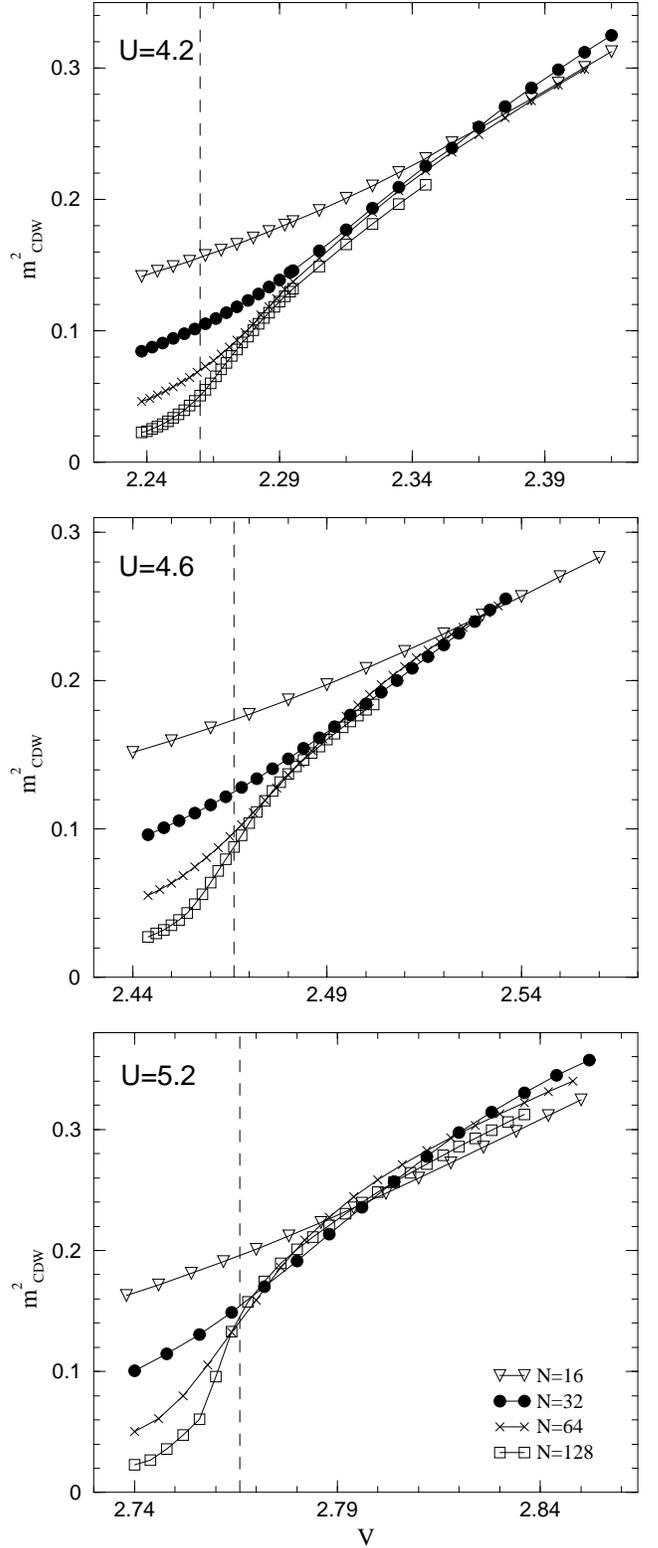}
\vskip1mm
\caption{CDW order parameter vs $V$ across the BOW-CDW boundary for 
several system sizes near the multi-critical point. The dashed line shows the 
position of $V_c$ for the respective $U$. Statistical errors are smaller
than the symbols.} 
\label{tricritical}
\end{figure}

By using larger system sizes and an alternative criterion to distinguish
between a continuous transition and a first-order level crossing 
transition, we have obtained an estimate of the multi-critical point that we
consider more accurate and reliable than the previous estimates. In contrast 
to most previous numerical studies, our method is not based on plotting 
histograms of the order parameter, although we will also present such 
histograms in the next section. In this section we first exploit the 
qualitatively different finite-size dependence of the growth of the order 
parameter close to the transition above and below the multi-critical point. 

For fixed $U$,
the order parameter curves for different system sizes cross each other at or 
very close to the critical point ($V=V_c$) in the case of a first-order 
transition, as discussed above in Sec.~III-D. Such a crossing cannot occur 
at a continuous transition, where instead there should be finite-size 
scaling governed by Eq.~(\ref{eq:fss}). This qualitative difference in the 
finite-size dependence of the order parameter close to the transition point 
above and below the multi-critical point $(U_m,V_m)$
leads us to expect that in the neighborhood of this point, curves of the 
order parameter for different chain lengths will closely coincide with one 
another close to $V=V_c$, and $U_m$ is the point at which the curves barely 
touch each other. When the system size becomes sufficiently large one can 
also directly observe discontinuities developing when $U > U_m$, in the order 
parameter as well as in other quantities, as in Fig.~\ref{u8}. In practice 
this criterion, or any other criterion known to us, cannot be expected to be 
useful very close to the multi-critical point, where the transition is only 
weakly first-order and very large lattices are needed to detect 
discontinuities developing from avoided level crossings. 

Fig.~\ref{tricritical} shows the finite-size dependence of the CDW order 
parameter across the transition for three different values of $U$.
For $U=4.2$, only the $N=16$ curve crosses the other curves, and this occurs 
far from the critical point (as determined using the peak in the charge 
stiffness, as discussed in Sec.~III-A). The non-crossing 
for larger system sizes show that the transition must be continuous at 
this $U$. For $U=5.2$ all curves show crossing behavior and a discontinuity 
can also be seen developing for the largest system size, i.e., the transition 
is here of first order. The curves for $U=4.6$ closely follow the expected 
behavior at the multi-critical point, with the curves for the largest systems 
barely touching each other. Based on data also for other values of $U$ we 
estimate the multi-critical point to be $(U_m=4.7\pm 0.1, V_m=2.51\pm 0.04)$. 
This agrees very well with Voit's estimate $(U_m=4.76)$.\cite{voit} 
However, it is not clear whether this agreement is fortuitous or whether 
there is some underlying symmetry that renders bosonization and RG (that 
assume $U,V \ll 1$) applicable close to the multi-critical point. 

\subsection{CDW order parameter histograms}

Previous studies of the multi-critical point have exploited the existence of 
a 3-peak structure in the distribution of the CDW order parameter for 
a discontinuous SDW-CDW transition in the vicinity of the critical point 
and its absence at a continuous transition.\cite{hirsch1} Outside the CDW 
phase, the distribution of the CDW order parameter is peaked around zero. 
For a continuous transition to a CDW state this peak splits into two 
(corresponding to positive and negative values of the order parameter), which 
gradually move apart from each other inside the CDW phase. In a first-order 
transition, on the other hand, the order parameter takes a non-zero value 
immediately as the CDW phase is entered and hence the two peaks emerge already
separated from each other. 
Furthermore, at the phase boundary the CDW phase coexists with 
the competing phase, and this is reflected as a central peak remaining in
the CDW order parameter distribution. The position of the multi-critical point
can then in principle be obtained by locating the point where the 3-peak
structure first appears. In practice, the accuracy of this method is limited 
by the fact that the discontinuity is very small for a first-order transition 
close to the multi-critical point and very large system sizes are then
needed to observe the three peaks. This problem is, of course, common to
all methods for distinguishing between a continuous and weakly first-order
transition.

\begin{figure}
\centering
\hskip-2mm
\epsfxsize=8.45cm
\leavevmode
\epsffile{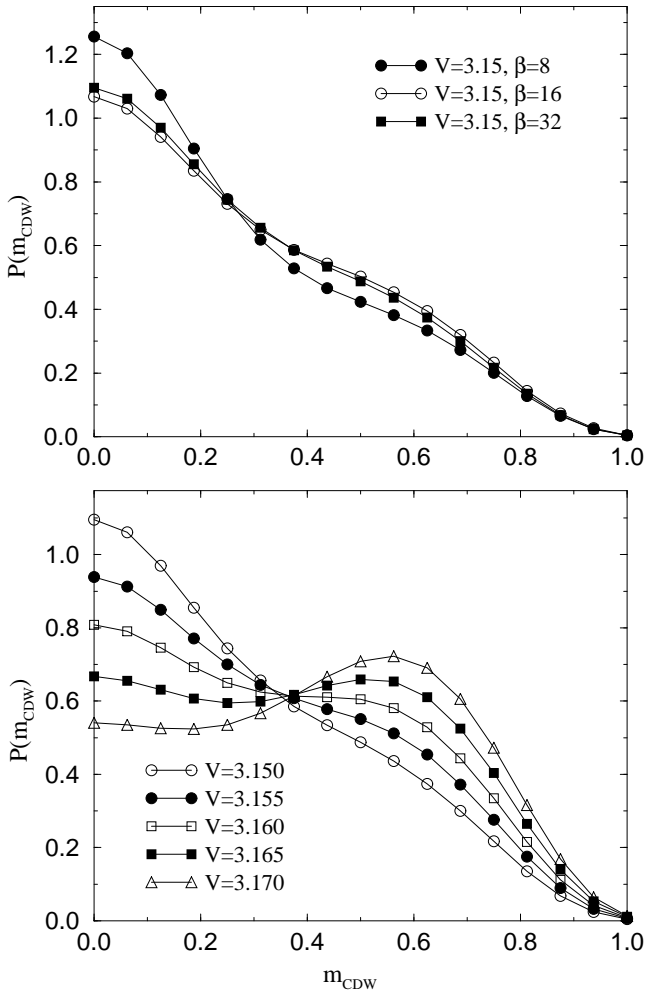}
\vskip1mm
\caption{CDW order parameter distributions for $32$-site systems at $U=6$. 
Upper panel: Dependence on the inverse temperature $\beta$ at $V=3.15$. Lower
panel: Dependence on $V$ around the first order phase transition. Statistical
errors are of the order of the size of the symbols.}
\label{hist32}
\end{figure}

In his early QMC study, Hirsch observed a 3-peak structure even for $U$ as
small as $3$ and therefore concluded that the transition is of first order
already there.\cite{hirsch1} For larger $U$, an unexplained $4$-peak structure
was seen. We have repeated histogram calculations for the lattice size $N=32$ 
studied by Hirsch. In Fig.~\ref{hist32} we show results for $U=6$, $V=3.15$,
where a 4-peak structure was seen in the earlier calculation.\cite{hirsch1} 
We only find a central peak, which show that the system is not in the 
CDW state for these parameters. There are, however, already signs of side
peaks developing, which shows that the system is close to the CDW phase.
The significant differences with the earlier result could partially be errors 
due to the Trotter decomposition used in the world-line simulation method. 
Temperature effects are only minor, as also shown in Fig.~\ref{hist32}. At 
$\beta=8$, which was used in Ref.~\onlinecite{hirsch1}, the histogram is only 
slightly more sharply peaked than at $\beta=16$ and $32$. Most likely, 
the simulation giving the $4$-peak structure was not sufficiently long, as
it consisted of only $10^4$ Monte Carlo steps.\cite{hirsch1} Even with the 
more efficient SSE algorithm used in the present work, we find that the 
autocorrelation times are quite long close to the first order transition 
(see Appendix B) and short simulation can produce incorrect order parameter 
histograms similar to those shown in Ref.~\onlinecite{hirsch1}. For the 
histograms shown here, on the order of $10^7-10^8$ SSE Monte Carlo steps 
were used.

\begin{figure}
\centering
\hskip-2mm
\epsfxsize=8.3cm
\leavevmode
\epsffile{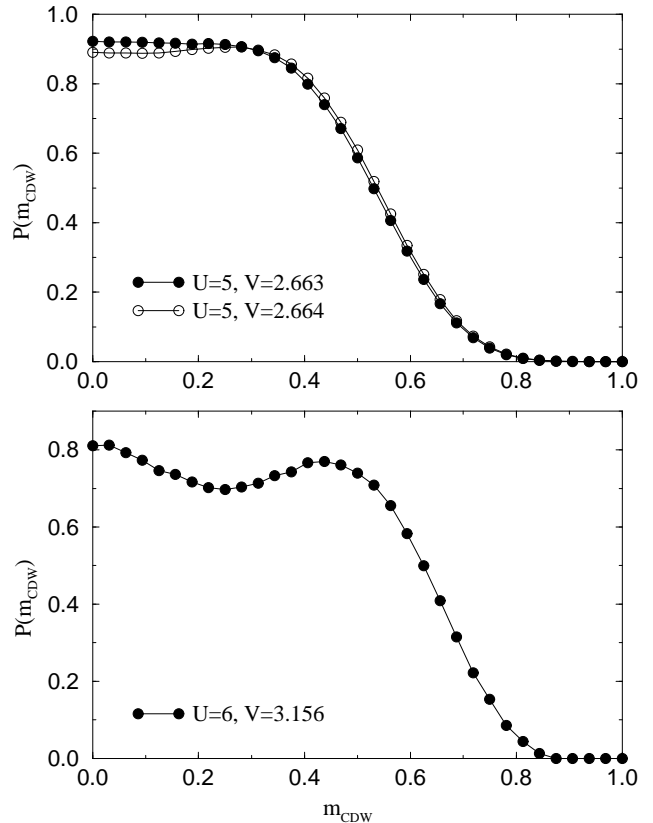}
\vskip1mm
\caption{CDW order parameter histograms for $N=64$ systems close to
the phase transition.}
\label{hist64}
\end{figure}

In Fig.~\ref{hist32} we also show results for several values of $V$ across
the phase transition. A clear 3-peak structure (i.e., three peaks in the range
$m_{CDW} \in [-1,1]$, of which we only show the positive part) with peaks of 
almost the same heights can be seen for $V=3.165$. In Fig.~\ref{hist64} we 
show results for $N=64$. At $U=6$, the 3-peak structure appears for 
$V \approx 3.156$, i.e, at a value slightly lower than for the $N=32$ system. 
The size of the $V$ region in which three peaks can be 
observed is also significantly smaller, reflecting the sharpening of the first
order transition cause by an avoided level crossing. At $U=5$, which we have 
argued above should be  close to but above the multi-critical point, we do
not observe three peaks. The histogram becomes very flat for an extended range
of $m_{CDW}$, however, and the side peak emerges at a finite value of 
$m_{CDW}$. This is consistent with the transition still being of first order 
at $V=5$. Going to still lower $V$-values, the peak just becomes narrower, 
and it is not possible to definitely conclude this way when the transition 
becomes continuous.

\section{Doped systems}

An interesting question naturally arises from the existence of the Nakamura 
BOW phase: Can the EHM model support a soliton lattice when doped slightly 
away from half-filling? Such a state exists in the  Su-Schrieffer-Heeger 
model in the adiabatic limit (with classical, ``frozen'' phonons),
\cite{ssh,kivelson,jeckelmann} and also when electron-electron 
interactions are taken into account.\cite{kivelson,jeckelmann} In these 
models, the quantum nature of the phonon field is not taken into account, 
however. It is known that the dimerized state at half-filling survives 
even in the presence of quantum fluctuations, at least up to a critical
value of the phonon frequency.\cite{fradkin} However, to our knowledge, 
there have been no reliable numerical calculations addressing the stability 
of the soliton lattice in the presence of fully quantum mechanical phonons. 
The Nakamura BOW is similar to a dimerized lattice with quantum fluctuations, 
and hence a study of its evolution with hole doping can give insights also 
into the quantum phonon problem. There are also unresolved issues regarding 
the doped CDW state.\cite{lin} In order to investigate the evolution of the 
long-range ordered states upon doping, we have studied the EHM model also 
away from half-filling, focusing on two parameter values in which the 
half-filled system is in the BOW (using $U=4$, $V=2.14$) or CDW phase 
(using $U=4$, $V=2.5$).

\begin{figure}
\centering
\epsfxsize=8.3cm
\leavevmode
\epsffile{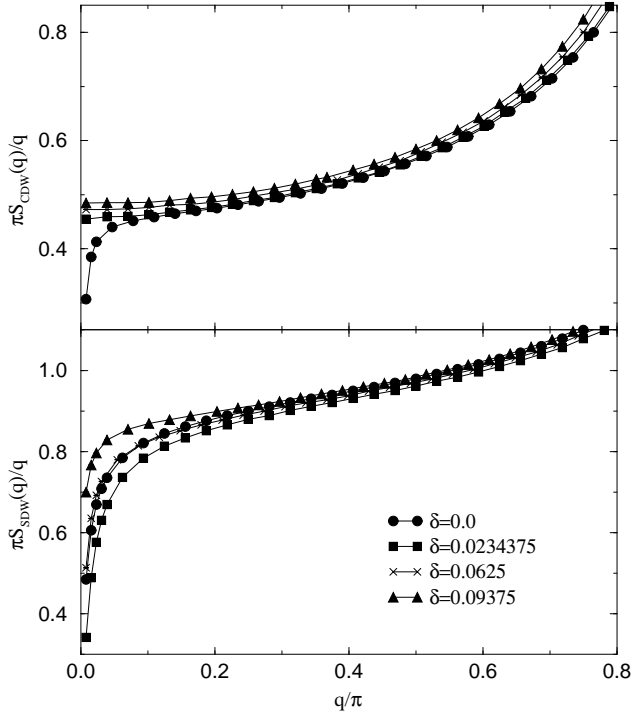}
\vskip1mm
\caption{The static charge (upper panel) and spin (lower panel) structure 
factor divided by the wave-number $q$ as a function of $q$ for $256$-site
chains at different doping fractions $\delta$. For these parameter values 
($U=4, V=2.14$), the half-filled system ($\delta=0$) is in the BOW state.}
\label{fig:bowsq}
\end{figure}

\begin{figure}
\centering
\epsfxsize=8.3cm
\leavevmode
\epsffile{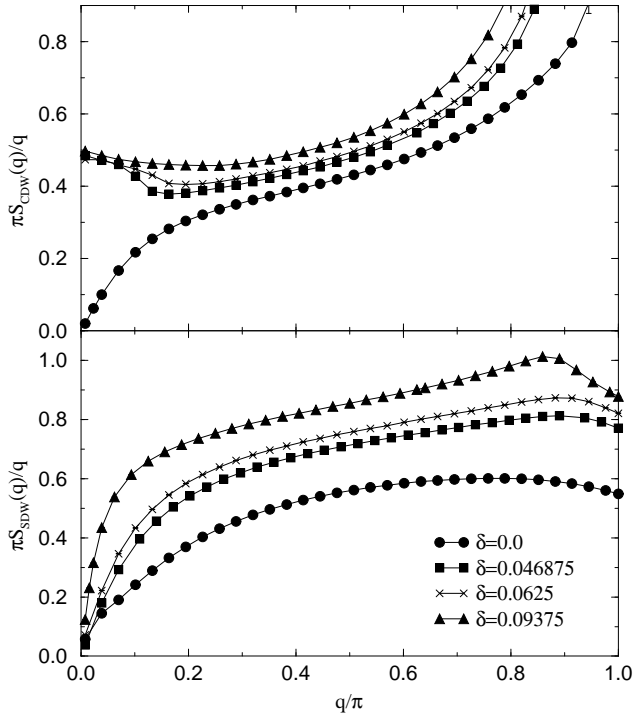}
\vskip1mm
\caption{Same as Fig.~\protect{\ref{fig:bowsq}} for $U=4, V=2.5$, where
the half-filled system is in the CDW phase.}
\label{fig:cdwsq}
\end{figure}

We first discuss the effects of doping on the spin and charge gaps
in the half-filled CDW and BOW ground states. As in previous sections, 
we make use of the behavior of the static structure factor in the limit 
$q\rightarrow 0$. Figs.~\ref{fig:bowsq} and \ref{fig:cdwsq} show 
$\pi S_{CDW}(q)/q$ and $\pi S_{SDW}(q)/q$ as a function of $q$ for a range 
of doping levels, both for parameters where the half-filled system is in the 
BOW phase (Fig.~\ref{fig:bowsq}) and in the CDW phase (Fig.~\ref{fig:cdwsq}).
From the data we conclude that upon doping away from half-filling, the charge 
gap vanishes immediately whereas the spin gap survives.
This is true for both the CDW and BOW parent states. This 
behavior is characteristic of a Luther-Emery liquid, \cite{le} 
in which the charge sector can be described in terms of a Luttinger liquid 
and the spin sector is gapped. The limiting value of $\pi S_c(q)/q$ as 
$q\rightarrow 0$ indicates that the Luttinger liquid exponent $K_{\rho}\approx
0.5$ in both the cases, with only a weak dependence on the doping level 
for the parameters considered here. Note the cross-over behavior occurring in
the charge structure at $q \approx 2\pi\delta =4k_F$ in Fig.~\ref{fig:cdwsq}
(which is accompanied by a peak in the corresponding susceptibility, as will 
be shown below),\cite{note4kf} reflecting a weak repulsion between dopant 
holes. No cross-over in the charge structure is seen in Fig.~\ref{fig:bowsq},
where the parent state is a BOW.

\begin{figure}
\centering
\epsfxsize=8.3cm
\leavevmode
\epsffile{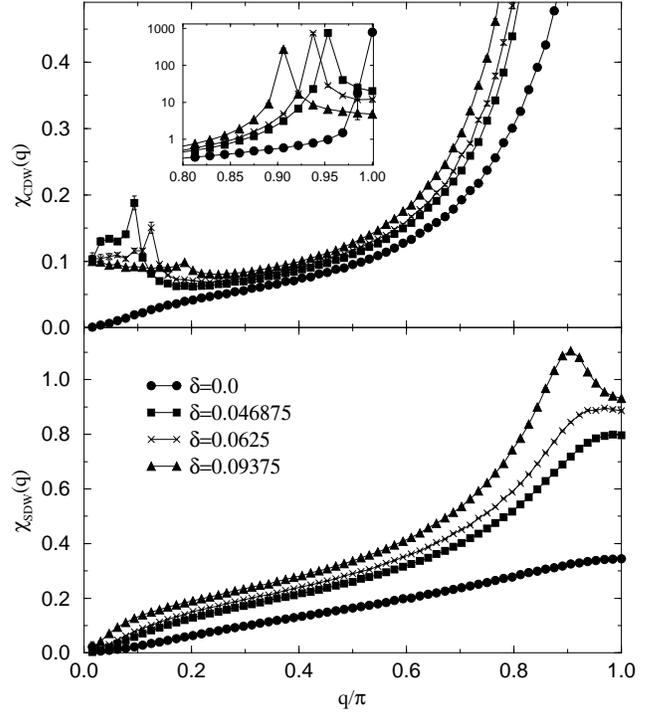}
\vskip1mm
\caption{Static CDW and SDW susceptibilities at different doping levels 
for $U=4$, $V=2.5$ for a $N=256$ chain. The inset shows the 
$\chi_{CDW}(2k_F)$ peaks on a more detailed scale.}
\label{fig:cdwxq}
\end{figure}

\begin{figure}
\centering
\epsfxsize=8.3cm
\leavevmode
\epsffile{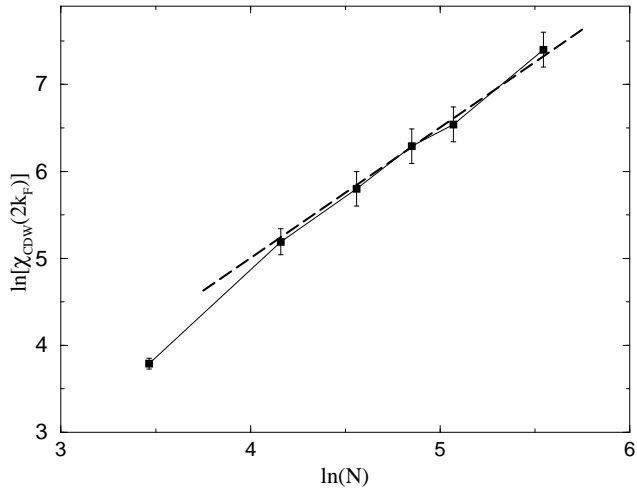}
\vskip1mm
\caption{Finite-size scaling of the static charge susceptibility at
$q=2k_F$ for a systems with $U=4,V=2.5$ at a doping level of 6.25\%. 
A slope of $1.5$ is shown by the dashed line.} 
\label{fig:lnx}
\end{figure}

Fig.~\ref{fig:cdwxq} shows the variation of the ground state static 
susceptibilities for several doping levels in a chain of length $N=128$ for 
the parameters $U=4,V=2.5$. For $\delta > 0$ the charge susceptibility 
converges to a non-zero value as $q\rightarrow 0$, again showing the absence 
of a charge gap. Very strong $2k_F$ peaks are evident, and weaker $4k_F$ 
peaks are also clearly visible. The $2k_F$ peaks diverge with the system 
size whereas the $4k_F$ peaks are non-divergent, in accord with the 
Luther-Emery picture. For a Luther-Emery liquid, the charge correlations
decay with distance $r$ as $r^{-K_\rho}$,\cite{voit} which gives 
$\chi_{CDW}(2k_F) \sim N^{2-K_{\rho}}$ for the finite-size scaling of the 
corresponding $2k_F$ susceptibility. Fig.~\ref{fig:lnx} shows the size 
dependence for $\delta=0.0625$ on a log-log scale. For system sizes 
$N\geq 64$, the data is seen to fall on a line with slope $\approx 1.5$, 
consistent with the value $K_\rho \approx 0.5$ extracted above.

\begin{figure}
\centering
\epsfxsize=8.3cm
\leavevmode
\epsffile{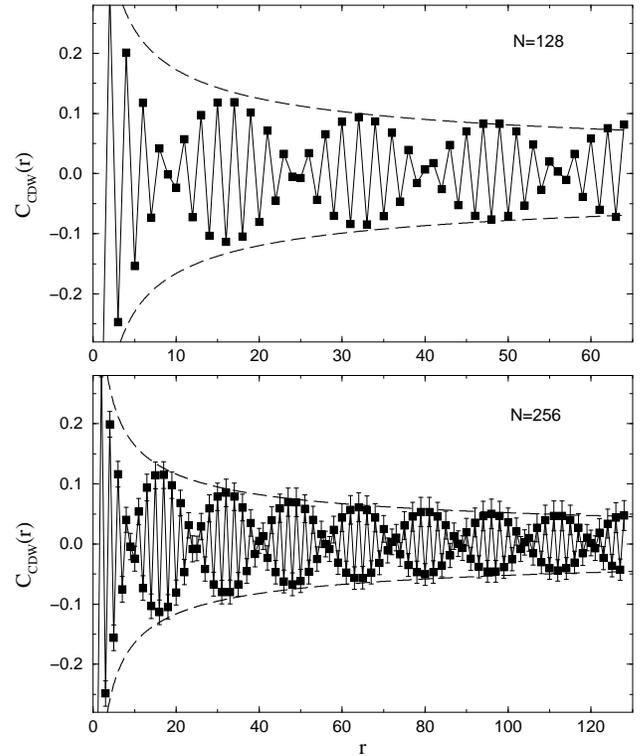}
\vskip1mm
\caption{Real-space charge correlations vs distance at a doping level
$\delta=0.0625$ fo system sizes $N=128$ and $256$ at $U=4,V=2.5$.} 
\label{fig:cr}
\end{figure}

As a further test of the Luther-Emery liquid nature of the ground
state away from half-filling, we have studied the real-space charge 
and bond correlations as a function of distance. Fig.~\ref{fig:cr} shows
the charge correlation for two different system sizes at a dopant 
concentration of 6.25\% and interaction parameters $U=4$ and $V=2.5$. 
The ground state at half-filling is a CDW, and for the doped system we find
solitonic features with alternating A and B phases separated by domain
walls or kinks. The correlation decay with distance, however, and there
is no real soliton lattice. In fact, the decay of the magnitude of the 
peaks is well approximated by an envelope curve of the form $y\sim x^{-0.5}$,
and hence also these data are consistent with a Luther-Emery state
with $K_{\rho}\approx 0.5$.

\begin{figure}
\centering
\epsfxsize=8.3cm
\leavevmode
\epsffile{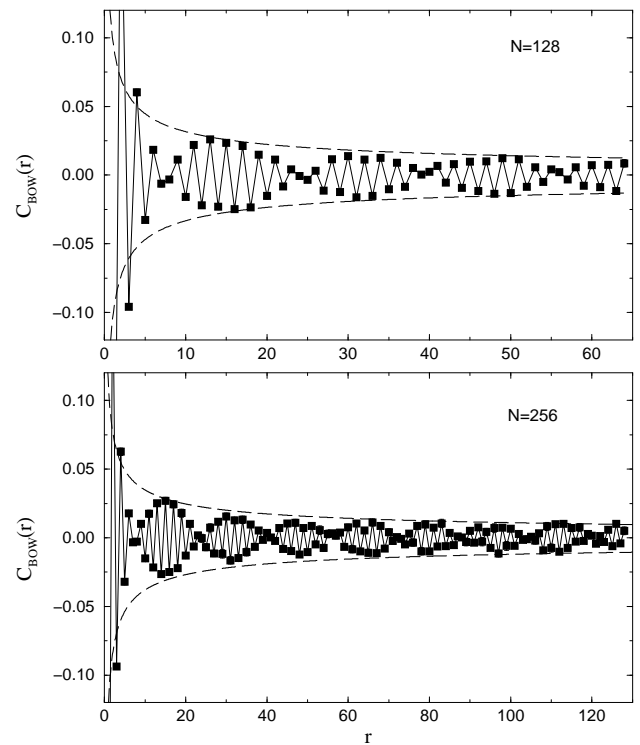}
\vskip1mm
\caption{Bond-order correlations at $\delta=0.0625$, $U=4, V=2.14$, for 
system sizes $N=128$ and $256$.} 
\label{fig:br}
\end{figure}

Fig.~\ref{fig:br} shows a similar plot of the real-space bond-order 
correlation for $U=4$ and $V=2.14$ at the same dopant concentration 
and for the same system sizes. The ground state of the half-filled
system for this choice of parameters is here a BOW, and away from 
half-filling, the dominant correlation are still of bond-order type. 
Once again, the ground state of the doped system has solitonic features 
with an algebraic decay of the magnitude of the peaks. As in the previous 
case, the decay is consistent with a Luther-Emery liquid with 
$K_{\rho}\approx 0.5$. 

\section{Summary}
\label{sec:summary}

To summarize, we have studied the 1D EHM using the SSE method incorporating
an efficient operator-loop update and a ``quantum parallel tempering'' scheme.
Our results 
confirm the surprising prediction \cite{nakamura} of the existence of a novel 
long-range-ordered BOW phase between the well-known CDW and SDW phases in the 
ground state phase diagram for small to intermediate values of the on-site 
interaction $U$ ($U<U_m$). We have presented several ways to detect the spin 
and charge gaps expected in the BOW phase and have also probed directly the 
BOW correlations and concluded that true long-range order develops. We have
studied a few points on the BOW-CDW phase boundary and obtained a very 
good agreement with Nakamura's level crossing prediction \cite{nakamura} 
for the location of this phase boundary. For the SDW-BOW phase boundary, our 
results indicate a higher critical $V$ for fixed $U$ than given by the level 
crossing method and thus over-all a slightly smaller size of 
the BOW phase. Our results are for significantly larger systems than in
the previous study and it is not surprising that the finite-size effects in 
the level crossings can be large for the SDW-BOW transition since the spin gap 
opens exponentially slowly in this Kosterlitz-Thouless transition. An 
over-estimation of the size of the BOW phase from the level crossings is also 
apparent considering that our estimated multi-critical point is well within 
the BOW phase of Nakamura's phase digram. Since our BOW-CDW phase boundaries 
agree, this indicates problems with the scaling of the exact SDW-BOW level 
crossings close to the multi-critical point, as was also mentioned by 
Nakamura.\cite{nakamura}
For large values of $U$ ($U>U_m$) the transition is  discontinuous 
(first-order). We have shown that curves of the CDW order parameter across 
this boundary for different system sizes cross each other twice, and explained
this behavior in terms of an avoided level crossing. We have also used the 
curve crossings as a means to locate the position of the multi-critical 
point with greater accuracy than previously attained. Our estimate 
for the multi-critical point is $U_m=4.7\pm 0.1, V_m=2.51 \pm 0.04$.

We have also studied systems doped slightly away from half-filling. We
find that both the doped CDW and BOW states give rise to ground states
of the Luther-Emery type, i.e., the quantum fluctuations do not allow
the formation of true soliton lattices. Based on the fact that the BOW
state is very similar to the dimerized ground state of models with
finite-frequency (non-adiabatic) phonons, we conjecture that the soliton
lattice is also unstable to arbitrarily weak quantum fluctuations in 
these models, unless two- or three-dimensional couplings are taken 
into account.

After our completion of the numerical calculations at half-filling 
Tsuchiizu and Furusaki \cite{tsuchiizu} presented a 
weak-coupling g-ology calculation taking 
into account second-order corrections to the coupling constants. They 
obtained a phase diagram in very good quantitative agreement with ours, 
including the location of the multi-critical point.

\acknowledgments

We would like to thank R. T. Clay for discussions of the operator-loop update 
for fermions. We thank M. Nakamura for sending us some of his numerical
results for comparisons. A.W.S. would like to thank O. Sushkov for 
discussions. This work was supported by the NSF under grant No.~DMR-97-12765
and by the Academy of Finland (project 26175). Most of the numerical 
calculations were carried out on the SGI Origin2000 and Condor systems at the 
NCSA, Urbana, Illinois. Some simulations were also carried out on the 
Origin2000 at CSC - Scientific Computing Ltd.~in Helsinki.

\appendix

\section{Operator-loop updates in the SSE method.}

The basic SSE approach has been discussed in several papers. 
\cite{anders1,anders2,sseloop}. We here start with a brief review as a basis 
for introducing the operator-loop update \cite{sseloop} in the context 
of fermion models.

To implement the SSE method, the Hamiltonian (\ref{H}) is written, upto an 
additive constant, in the form 
\begin{equation}
H=-\sum_{b=1}^{N}(H_{1,b} + H_{2,b} + H_{3,b}),
\end{equation}
where $b$ is the bond connecting the sites $b$ and $b+1$, $N$ is the length 
of the chain, and the operators $H_{a,b}, a=1,2,3$ are defined as
\begin{eqnarray}
H_{1,b} &=&
C-\hbox{$U\over 2$}(n_{b,\uparrow}-\hbox{$1\over 2$}) 
(n_{b,\downarrow}-\hbox{$1\over 2$}) \nonumber \\ 
& - & \hbox{$U\over 2$}(n_{b+1,\uparrow}-\hbox{$1\over 2$})
(n_{b+1,\downarrow}-\hbox{$1\over 2$}) \nonumber \\
	& - & V(n_b-1)(n_{b+1}-1),\\
H_{2,b} &=& t(c^{\dagger}_{b+1,\downarrow}c_{b,\downarrow} + h.c.), 
\nonumber \\
H_{3,b} &=& t(c^{\dagger}_{b+1,\uparrow}c_{b,\uparrow} + h.c.). 
\nonumber
\end{eqnarray}
The constant $C$ shifts the zero of the energy and is chosen to 
ensure a non-negative expectation value for $H_{1,b}$ (needed in order to
ensure a positive definite expansion of the partition function). Introducing 
a basis
$\{|\alpha\rangle\}=\{|\zeta_1,\zeta_2,\dots,
\zeta_N\rangle\}$, where $\zeta_i\in\{0,\uparrow,\downarrow,
\uparrow\downarrow\}$ denotes the electron state at the site
$i$, the partition function $Z$=Tr$\{e^{-\beta H}\}$ can be expanded in a
Taylor series as
\begin{equation}
Z=\sum_{\alpha}\sum_{n=0}^{\infty}\sum_{S_n}\frac{\beta^n}{n!}
   \left \langle\alpha \left |\prod_{p=1}^n H_{a_p,b_p}\right |\alpha
    \right\rangle,
\label{Zn}
\end{equation}
where $S_n$ denotes a sequence of index pairs defining the operator
string $\prod_{p=1}^n H_{a_p,b_p}$:
\begin{equation}
S_n=[a,b]_1[a,b]_2\dots [a,b]_n,
\label{sn}
\end{equation}
where we use the notation $[a,b]_p = [a_p,b_p]$ and
$a\in\{1,2,3\}$, $b\in\{1,\dots,N\}$.
In order to construct an 
efficient updating scheme, the Taylor series is truncated at a 
self-consistently determined power $L$, large enough to cause only an 
exponentially small, completely negligible error ($L \sim \beta |E|$, where 
$E$ is the total internal energy; for details see 
Refs.~\onlinecite{anders1,anders2}). 
We can then define a sampling space where the length of the sequences is
fixed, by inserting $L-n$ unit operators, denoted by $H_{0,0}$, into each 
sequence. The terms in the partition function must be divided by $L\choose n$ 
in order to compensate for the different ways of inserting the unit operators. 
The summation over $n$ is then implicitly included in the summation over 
all sequences of length $L$. The partition function takes the form
\begin{equation}
Z=\sum_{\alpha}\sum_{S_L}\frac{\beta^n(L-n)!}{L!} 
   \left\langle\alpha\left|\prod_{p=1}^L H_{a_p,b_p}\right|\alpha\right\rangle,
\label{Z}
\end{equation}
where the operator-index pairs $[a,b]_p$ now have
$a\in\{1,2,3\}$ and $b\in\{1,\dots,N\}$ or $[a,b]_p=[0,0]$. For convenience, 
we introduce a notation for states obtained by the action of the first $p$ 
elements of the operator string $S_L$:
\begin{equation}
|\alpha(p)\rangle \sim \prod_{j=1}^p H_{a_j,b_j}|\alpha\rangle.
\end{equation}
For a nonzero contribution to the partition function, $|\alpha(L)\rangle = 
|\alpha(0)\rangle$. 

A Monte Carlo scheme is used to sample the configurations $(\alpha,S_L)$ 
according to their relative contributions (weights) to $Z$. The sampling 
scheme consists of two types of updates,\cite{anders1,anders2,sseloop} 
referred to as diagonal update and operator-loop updates. The diagonal update 
involves local substitutions of the form $[0,0]_p \leftrightarrow [1,b]_p$ and 
is attempted consecutively for every $p\in\{1,\dots,L\}$ in the sequence for 
which $[a,b]_p=[0,0]_p$ or $[1,b]_p$. The updates are accepted with 
probabilities 
\begin{eqnarray}
P([0,0]_p\rightarrow [1,b]_p) &=& \
\frac{N\beta M_{1,b} (p)}{L-n},
\nonumber \\
P([1,b]_p\rightarrow [0,0]_p) &=& \
\frac{L-n+1}{N\beta M_{1,b}(p)}, 
\label{eq:pdiagonal}
\end{eqnarray}
where 
\begin{equation}
M_{a,b}(p) = \langle \zeta_b(p),\zeta_{b+1}(p) |H_{a,b}
|\zeta_b(p-1),\zeta_{b+1}(p-1)\rangle
\label{eq:melement}
\end{equation}
is a matrix element on bond $b$, which in this case is diagonal ($a=1$). 
Only a single state $|\alpha (p)\rangle$ is stored in the computer during the 
diagonal update. When off-diagonal operators are encountered during the 
successive scanning of the operator string, the corresponding electron states 
are updated so that the information needed for evaluation of the probabilities 
(\ref{eq:pdiagonal}) is always available when needed.

\begin{figure}
\centering
\epsfxsize=8.3cm
\leavevmode
\epsffile{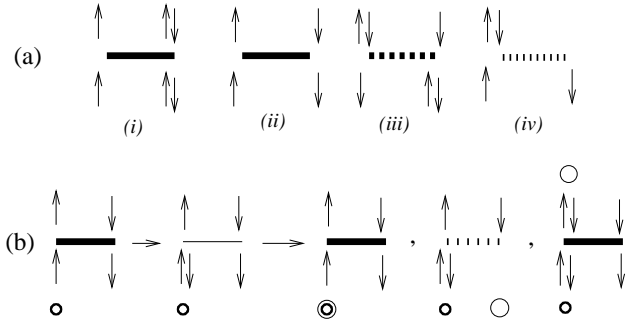}
\vskip2mm
\caption{(a) A few allowed vertices. The solid lines denote the diagonal
Hamiltonian operator, the dashed and dotted lines denote the hopping operators
for the up and down spins respectively. The lower legs denote the states
$\zeta_i(p-1)$ and $\zeta_{i+1}(p-1)$ while the upper legs denote $\zeta_i(p)$ 
and $\zeta_{i+1}(p)$. (b) An example of a vertex update. The entrance leg is 
the lower left leg of the vertex, as indicated by the dot. The electron state 
at the entrance leg, $\uparrow$, is changed to $\uparrow\downarrow$ in this
particular update. Given that, the 3 possible resulting vertices are shown. 
The corresponding exit legs are denoted by open circles. Exit at the upper 
right leg does not result in an allowed vertex in this case.} 
\label{vx}
\end{figure}

The operator-loop update has been discussed in detail in 
Ref.~\onlinecite{sseloop} in the context of spins. Here we present the 
construction of loops for fermions. As explained in Ref.~\onlinecite{sseloop}, 
the matrix element in Eq.~(\ref{Z}) can be graphically represented by a set of 
$n$ vertices (corresponding to the $n$ non-unit operators in $S_L$) connected 
to one another by the propagated electron states. Each vertex has four ``legs'' 
with electron states $|\zeta_i(p-1),\zeta_{i+1}(p-1)
\rangle$ and $|\zeta_i(p),\zeta_{i+1}(p)\rangle$ before and after the action 
of the associated Hamiltonian operator $H_{a_p,b_p}$. There are 32 allowed
vertices -- 16 diagonal ones and 8 each associated with the off-diagonal
$H_{2,b}$ and $H_{3,b}$ [see Fig.~\ref{vx}(a)].
 A configuration $(\alpha, S_L)$ is completely 
specified by the leg states of the $n$ vertices --- except for sites
that do not have any operators acting on them.

To carry out the operator-loop update, the linked list of the $n$ vertices is 
first constructed. In addition to the electron states at the legs of each 
vertex, the list also contains the addresses (i.e., the location in $S_L$) of 
the next vertex and the corresponding leg that each leg is connected to. The 
loop construction begins with randomly choosing a vertex and an ``entry'' leg.
The electron state at the entry leg is changed to one of the 3 other allowed 
states chosen at random. Next an ``exit'' leg is chosen (following a procedure 
described below) and its associated electron state is updated so that the new 
leg states constitute an allowed vertex [see Fig.~\ref{vx}(b)]. The exit leg 
will be linked to a leg 
of another vertex (or, if there is only one operator in the configuration 
which 
acts on the site in question, another leg on the same vertex) and this will be 
the entry leg for the next vertex. The electron state at this new entry leg 
is then updated to match the state at the exit leg of the previous vertex. 
A new exit leg is then chosen following the same procedure. This is repeated 
until the exit leg from a vertex points to the starting point of the loop, 
which implies that the loop is closed and a new allowed configuration has 
been generated.

To choose an exit leg --- given a vertex, an entry leg and the updated 
electron state at the entry leg --- all the legs can be considered in turn 
and attempts made to update the associated electron state so that the new 
leg states constitute an allowed vertex. Because of spin an charge 
conservation on the vertices, at a given exit leg there is at most one 
possible update of the electron state that can lead to an allowed vertex. 
Hence, the exit leg uniquely determines the new vertex and the probability
of choosing a given leg should be proportional to the weight of the new 
vertex, i.e., a matrix element of the form (\ref{eq:melement}), which in
this case can be either diagonal or off-diagonal. 
In practice, a fast selection of an exit leg and 
updating of the vertex state is achieved using two pre-generated tables. 
The first one contains the cumulative probabilities of the 
4 exit legs given an entrance leg, the old vertex state, and the new state at 
the entrance. The second table contains the new vertex states corresponding to 
the updated entrance and exit legs.

A special case occurs if the initial update at the entry leg of the first 
vertex of a loop is a spin-flip, i.e. the electron state changes from 
$\uparrow$ to $\downarrow$ or vice versa. In this case, the vertex weight
does not change when updated and as a consequence the ``bounce process'', 
where the exit leg is the same as the entrance leg, does not have to be
included in the loop construction. The loop then becomes deterministic,
i.e., there is a unique exit leg given the entrance leg.\cite{sseloop}
This is similar to the ``loop-exchange'' algorithm proposed in the
context of the world-line method.\cite{kawashima}

A full Monte Carlo updating cycle (MC step) consists of a diagonal update, 
followed by the construction of a linked vertex list. Next a number of 
operator-loop updates are carried out and finally the vertices are mapped 
back into a corresponding sequence $S_L$. The loop update typically also 
implies changes in the stored state $|\alpha\rangle = |\alpha (0) \rangle$, 
as some of the vertex legs (links) span across the periodic boundary in the 
propagation direction. The number of up and down electrons can be changed by
the operator-loop update, as can the spatial winding numbers, and the algorithm is hence fully grand canonical. Note that at high and moderately 
low temperatures there are typically some sites of the system which has no 
vertices associated with them. The states on these sites can be randomly 
changed, since they have no affect the configuration weight.

The number of loops constructed for every MC step is determined such that on 
an average a total of $\sim L$ vertices (we typically use $2L$) are visited. 
The truncation $L$ and the number of loops are adjusted during he 
equilibration 
part of the simulation and are thereafter held fixed. $L$ is determined by
requiring that the highest $n$ reached during equilibration is at most
$70-80\%$ of $L$.

In certain parameter regions the length of a loop can sometimes become 
extremely 
long before it closes --- in practice, it may even never close. 
It is therefore necessary 
to impose a maximum length, beyond which the loop construction is terminated 
and a new starting point is chosen (typically, we use $\approx 50L$ 
for this cut-off length). In order to reduce the likelihood of the next 
loop also exceeding the termination length it has proven useful 
to carry out a diagonal update before starting 
the next loop. The loop termination does not violate detailed balance and does 
not cause any systematical errors in the results. In most cases, incomplete
loop termination occurs so infrequently that it does not adversely affect the 
simulation. In analogy with Ref.~\onlinecite{prokofev}, where a scheme
(there called ``worm'' update) similar to the operator-loops considered 
here was first introduced within the continuous worl-line representation,
the end points of the loop during construction can be related to the 
single-particle Green's function of the system and hence the tendency for 
loops to become exceedingly long for some parameter values must be related 
to some physical properties of the system. This issue should be studied 
further.

Estimators for the various structure factors and susceptibilities have
been discussed in previous articles.\cite{anders1,anders2} Here we only note
that the charge and spin 
stiffness constants, Eq.~(\ref{eq:rhoc}), can be expressed in terms of spin 
and charge current operators in analogy with the spin stiffness of the 
Heisenberg antiferromagnet previously discussed in Ref.~\onlinecite{anders2},
leading to 
\begin{equation}
\rho_{c,s}=\frac{[(n_R^\uparrow-n_L^\uparrow) 
             \pm  (n_R^\downarrow-n_L^\downarrow)]^2 }{N\beta},
\end{equation}
where $n_{R,L}^{\sigma}$ are the number of kinetic energy operators in 
the SSE term propagating spin-$\sigma$ particles in the ``right'' and ``left'' 
direction on the ring. Because of spin and charge conservation, the 
topological winding numbers $(n_R^\sigma-n_L^\sigma)/N$ can take only 
integer values. 

\begin{figure}
\centering
\epsfxsize=8.3cm
\leavevmode
\epsffile{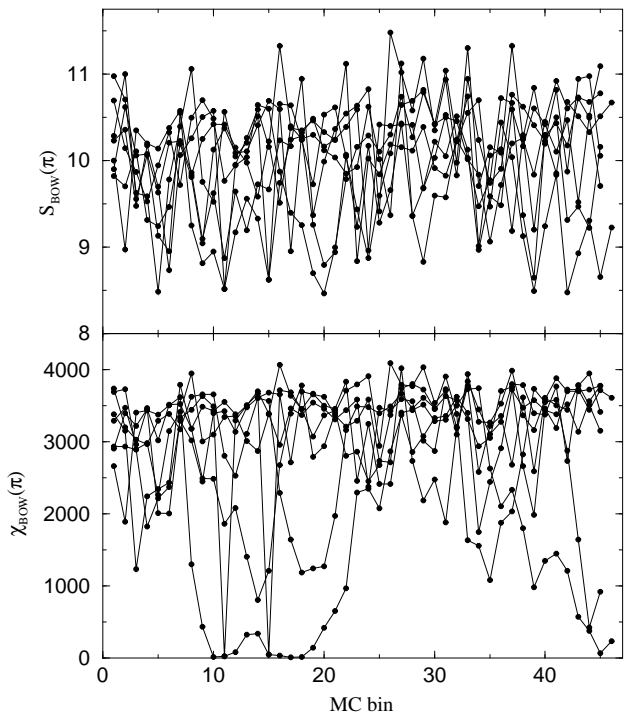}
\vskip1mm
\caption{The BOW structure factor and susceptibility for a $256$-site system
with $U=4$ and $V=2.14$ at inverse temperature $\beta=512$. Results of six
independent simulations are shown. Each point represents an average over
a bin consisting of $10^4$ Monte Carlo steps.} 
\label{fig:dynamics}
\end{figure}

Although the operator-loop algorithm very significantly speeds up SSE 
simulations, in many cases reducing the autocorrelation function by orders of 
magnitude, the dynamics is still very slow in some parameter regions. For the 
extended Hubbard model studied here, problems with very long autocorrelation 
times occur in the long-range ordered BOW and CDW phases. The problems are 
particularly severe for large systems close to the BOW-CDW phase 
boundary, where ``trapping''in the wrong phase often occurs. The slow dynamics
in the BOW phase is illustrated in Figure \ref{fig:dynamics}, which shows
the simulation time dependence of $S_{BOW}(\pi)$ and $\chi_{BOW}(\pi)$ during 
a simulation of a $256$-site system at $\beta=512$ [$S_{BOW}$ has 
converged at this $\beta$ but $\chi_{BOW}$ is about $20\%$ larger still at 
$\beta=1024$]. It is evident that the BOW autocorrelation time here is 
tens of thousands of MC steps. The BOW susceptibility exhibits a behavior 
where it sometimes takes very small values (less that $10^{-3}$ of the average
value), but corresponding large fluctuations upwards do not occur, i.e., the 
distribution of the $\chi_{BOW}(\pi)$ estimator for individual configurations
is very skewed. The structure factor exhibits a more symmetric distribution. 
This behavior can be understood as a consequence of the BOW ground state for 
a finite system being a symmetric combination of the two possible real-space 
symmetry-broken states. The symmetry is not broken in a finite system and the 
simulation is also not trapped in one of the real space-states. Hence, the 
wave function that is sampled in the simulation contains both the real-space 
states and the behavior seen in Fig.~\ref{fig:dynamics} indicates that 
individual configurations also contain both components, in such a way that 
transitions (``tunneling'') between the two real-space states can occur 
during the SSE propagation (which can be simply related \cite{irsse} to a 
propagation in imaginary time), at least for some configurations. Tunneling
can be inferred from the qualitatively different evolutions of the structure 
factor and the susceptibility in MC time. The susceptibility is an integral 
of the bond-order correlation, as in 
Eq.~(\ref{eq:xcdw}), which in configurations where tunneling occurs can be 
much smaller than in configurations with no tunneling, since correlations 
between states with the same real-space configuration contributes positively 
but correlations between different states give a negative contribution. The 
structure factor, on the other hand, is an equal time correlation function 
and would not be much reduced by tunneling if the tunneling times are short. 
This explains the qualitatively different distributions of the $\chi_{BOW}$ 
and $S_{BOW}$ measurements in Fig.~\ref{fig:dynamics}. Evidently, the updating 
process is very slow in adding and removing tunneling events in the 
configurations, which maybe is not that surprising considering that the 
tunneling is between two states with a discrete broken symmetry. These 
problems do not occur in SSE simulations of systems with a broken continuous 
symmetry, such as the two-dimensional Heisenberg model.

The trapping and tunneling problems can be significantly reduced by using
the parallel tempering scheme (or exchange Monte Carlo),
\cite{tempering1,tempering2,tempering3}
which is discussed below in Appendix B.

\section{Quantum parallel tempering}

The ``quantum parallel tempering'' scheme is a straight-forward generalization 
of the thermal parallel tempering \cite{tempering1,tempering2,tempering3} 
method commonly used to equilibrate classical spin glass simulations. Our 
implementation amounts to running several simulations simultaneously on a 
parallel computer, using a a fixed value of $U$ and different closely spaced 
values of $V$. Along with the usual Monte Carlo updates, we 
attempt to swap the configurations for processes with adjacent values of $V$ 
at regular intervals, typically after every Monte Carlo step, according to a 
scheme that maintains detailed balance in the extended ensemble of parallel
simulations. The probability of swapping the $V$-values of runs $i$ and $i+1$, 
which are running at $V_i$ and $V_{i+1}$, respectively, before the swap, is
\begin{equation}
P_{\rm swap}(V_i,V_{i+1})=\rm{min}
[1,\frac{W_i(V_{i+1})W_{i+1}(V_{i})}{W_i(V_{i})W_{i+1}(V_{i+1})}],
\label{eq:pswitch}
\end{equation}
where $W_{i}(V)$ is the weight of the $i$th simulation configuration evaluated 
with the coupling $V$. The swap probabilities for fixed 
$\Delta V = V_{i+1}-V_i$ decreases with increasing system size and decreasing 
temperature and hence $\Delta V$ and the range of $V$-values (if the number of 
processes is fixed) must be chosen smaller for larger system sizes.

The computational effort required for the swapping process is very minor 
compared to the actual quantum Monte Carlo simulations. It is therefore useful
to carry out several swap attempts of all pairs of neighboring simulations
between every MC step. Histograms containing the number of times each of the
current configurations has ``occupied'' each $V$-bin can then be constructed 
and used for adding the contributions of each configuration to all the 
$V$-bins. This can contribute to reducing the statistical error of measured 
quantities.

\begin{figure}
\centering
\epsfxsize=8.3cm
\leavevmode
\epsffile{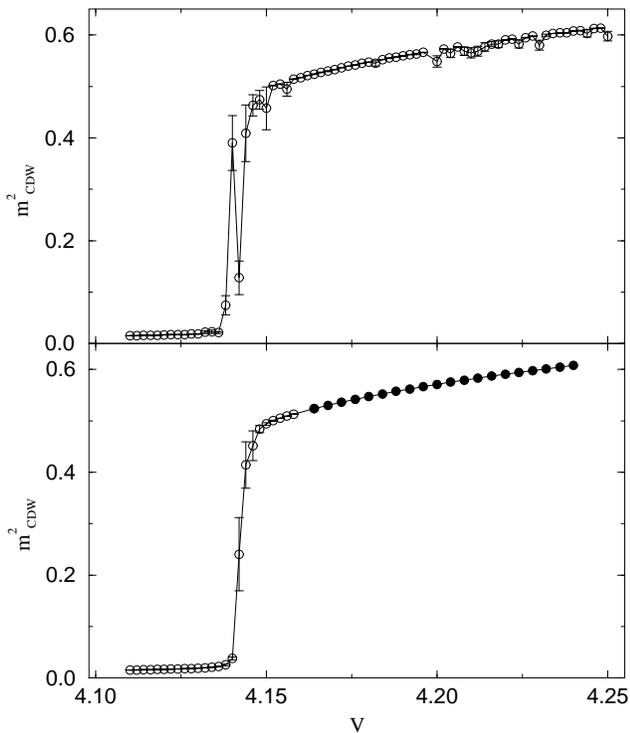}
\vskip1mm
\caption{The CDW order parameter across the SDW-CDW
phase boundary for $U=8$ ($N=64$, $\beta=64$). The upper panel shows data 
from individual runs. The lower panel shows the same data obtained using 
quantum parallel tempering, with two independent runs as indicated by the 
open and solid circles.} 
\label{fig:tempering}
\end{figure}

To illustrate the advantage of quantum parallel tempering, we show two 
sets of data --- obtained with and without the use of tempering --- for a 
system undergoing a first-order transition. Fig.~\ref{fig:tempering} shows the 
CDW order parameter across the first-order SDW-CDW phase boundary at $U=8$. 
The upper panel shows the data obtained from individual runs; 
the lower panel shows data for the same parameters obtained using tempering. 
The length of the individual simulations was $10^5$ MC steps for all $V$ 
values, and this was also the number of steps performed by each process in 
the tempering runs. The improvement in the quality of the tempering data
is evident, especially close to the transition point where two of the 
individual simulations have relaxed into the wrong phases. The statistical 
errors are hence severely underestimated due to the failure to equilibrate 
properly within the simulation time. The tempering error bars are also large 
at the phase transition, but in contrast to those of the individual 
simulations they are accurate error estimates. The errors 
rapidly become much smaller as one moves away from the transition point. 
The effects of tempering are also favorable further inside the CDW phase, 
where several of the individual simulations are apparently affected by 
trapping in configurations with defects, where the order is reduced.

\begin{figure}
\centering
\epsfxsize=8.3cm
\leavevmode
\epsffile{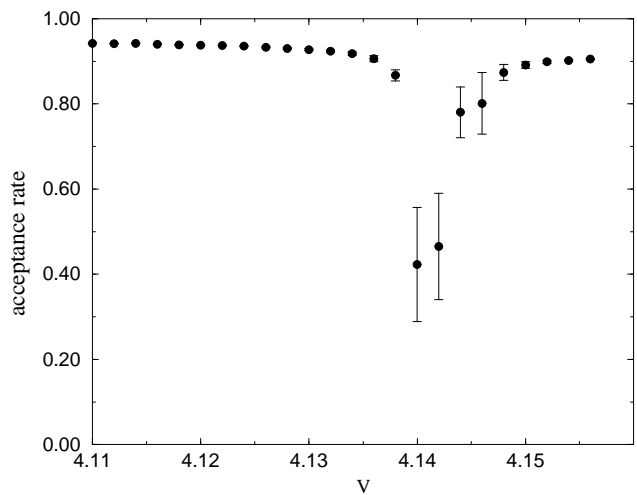}
\vskip1mm
\caption{The tempering acceptance rate during the simulation across the
first-order SDW-CDW phase boundary.}
\label{fig:accept}
\end{figure}

The tempering acceptance rate during the run spanning across the 
phase transition in Fig.~\ref{fig:tempering} is shown in Fig.~\ref{fig:accept}.
 There is a sharp reduction in the acceptance rate at the transition. This 
reflects the rapid change in the SSE configurations across the phase boundary, 
which implies that the configuration weights evaluated with $V$ values from 
the ``wrong'' phase are likely to decrease and the swap according to
the probability (\ref{eq:pswitch}) to be rejected.

Finally, we note that tempering, in general, is an application where a
superlinear speed-up can be achieved in practice on parallel computers.
In addition to doubling the density of data points when the number of
processes is doubled, the statistical errors are also reduced. Sometimes
the error reduction can be dramatic, but even in cases where there are
no real problems with the dynamics of individual simulations the effects 
of tempering are often very favorable. 

\null


\vskip-4mm
\begin{references}
\null\vskip-20mm\null
\bibitem{cuo}

V. J. Emery, S. A. Kivelson and O. Zachar, Phys Rev. B {\bf 56}, 6120 
(1997).

\bibitem{-ch-ch-}
H. G. Keiss (ed.~), {\it Conjugated Conducting Polymers}, 
(Springer-Verlag, Berlin, 1992).

\bibitem{ttftcnq}
T. Ishiguro and K. Yamaji, {\it Organic Superconductors}, 
(Springer-Verlag, Berlin, 1990).

\bibitem{liebwu}
E. H. Lieb and F. Y. Wu, Phys. Rev. Lett. {\bf 20} 1445 (1968).

\bibitem{bari}
R. A. Bari, Phys. Rev. B {\bf 3}, 2662 (1971).

\bibitem{emery1}
V. J. Emery, in {\it Highly Conducting One-Dimensional Solids}, 
ed. J. T. Devreese, R. Evrand and V. van Doren, (Plenum, New York, 1979), 
p.327.

\bibitem{solyom}
J. S\'{o}lyom, Adv. Phys. {\bf 28}, 201 (1979).

\bibitem{hirsch1}
J. E. Hirsch, Phys. Rev. Lett. {\bf 53}, 2327 (1984).

\bibitem{hirsch2}
J. E. Hirsch, Phys. Rev. B {\bf 31}, 6022 (1985).

\bibitem{fradkin1}
J. W. Cannon and E. Fradkin, Phys. Rev. B {\bf 41}, 9435 (1990).

\bibitem{fradkin2}
J. W. Cannon, R. T. Scalettar and E. Fradkin, Phys. Rev. B {\bf 44}, 
5995 (1991).

\bibitem{vandongen}
P. G. J. van Dongen, Phys. Rev. B {\bf 49}, 7904 (1994).

\bibitem{fourcade}
B. Fourcade and G. Spronken, Phys. Rev. B {\bf 29}, 5096 (1984).

\bibitem{voit}
J. Voit, Phys. Rev. B {\bf 45}, 4027 (1992).

\bibitem{lin}
H. Q. Lin, E. R. Gagliano, D. K. Campbell, E. H. Fradkin, and J. K.
Gubernatis, in {\it Proceedings of the 1993 NATO ARW on the Physics and
Mathematical Physics of the Hubbard model}, edited by D. Baeriswyl {\it et al.}
(Plenum, New York, 1995).

\bibitem{nakamura} 
M. Nakamura, J. Phys. Soc. of Japan {\bf 68}, 3123 (1999), 
Phys. Rev. B {\bf 61}, 16377 (2000).

\bibitem{anders1}
A. W. Sandvik and J. Kurkij{\"a}rvi, Phys. Rev. B {\bf 43}, 5950 (1991);
A. W. Sandvik, J. Phys. A {\bf 25}, 3667 (1992).

\bibitem{anders2}
A. W. Sandvik, Phys. Rev. B {\bf 56}, 11678 (1997).

\bibitem{sseloop}
A. W. Sandvik, Phys. Rev. B {\bf 59}, R14157 (1999).

\bibitem{tempering1}
K. Hukushima, H. Takayama, K. Nemoto, Int. J. Mod. Phys. C {\bf 7}, 
337 (1996)

\bibitem{tempering2}
K. Hukushima, K. Nemoto, J. Phys. Soc. Jpn. {\bf 65}, 1604 (1996)

\bibitem{tempering3}
E. Marinari, Lecture Notes in Physics, Vol.~501 {\it Advances in computer 
simulation: Lectures held at the E\"{o}tv\"{o}s Summer School, Budapest, 
Hungary, 16-20, July 1996}, edited by J. Kertsz and I. Kondor (Springer, 1998).

\bibitem{bilayer}
P. V. Shevchenko, A. W. Sandvik, and O. P. Sushkov,
Phys. Rev. B, {\bf 61}, 3475 (2000).

\bibitem{wessel}
S. Wessel, B. Normand, M. Sigrist, and S. Haas, Phys. Rev. Lett. {\bf 86},
1086 (2001).

\bibitem{dorneich}
A. Dorneich and M. Troyer, Phys. Rev. E {\bf 64}, 066701 (2001). 

\bibitem{kohn}
W. Kohn, Phys. Rev. {\bf 133}, A171 (1964).

\bibitem{periodicnote}
Strictly speaking, our boundary conditions are antiperiodic, which is
what one automatically obtains at half-filling when the system size $N$ 
is a multiple of 4 and the fermion anticommutation relations giving negative 
signs from transport across the boundary are neglected.

\bibitem{ll}
J. Voit, Rep. Prog. Phys. {\bf 57}, 977 (1994)

\bibitem{torsten}
R. T. Clay, A. W. Sandvik and D. K. Campbell, Phys. Rev. B {\bf 59}, 
4665 (1999).

\bibitem{schulz}
H. J. Schulz, Phys. Rev. Lett. {\bf 64}, 2831 (1990); Int. J. Mod. Phys.
{\bf 5}, 57 (1991).

\bibitem{eckle}
F. Woynarovich and H.-P. Eckle, J. Phys. A {\bf 20}, L97 (1987). 

\bibitem{eggert1}
S. Eggert, I. Affleck, and M. Takahashi, Phys. Rev. Lett. {\bf 73}, 332 (1994).

\bibitem{eggert2}
S. Eggert, Phys. Rev. B {\bf 54}, R9612 (1996).  

\bibitem{ktnote}
H. Weber and P. Minnhagen, Phys. Rev. B {\bf 37}, 5986 (1987); 
K. Harada and N. Kawashima, J. Phys. Soc. Jpn. {\bf 67}, 2768 (1998).

\bibitem{sppaper}
A. W. Sandvik and D. K. Campbell, Phys. Rev. Lett. {\bf 83}, 195 (1999).

\bibitem{nomura}
K. Nomura and K. Okamoto, J. Phys. A {\bf 27}, 5773 (1994);
J. L. Cardy, J. Phys. A {\bf 20}, L891 (1987).

\bibitem{ssh}
W. P. Su, J. R. Schrieffer, and A. J. Heeger, Phys. Rev. Lett. {\bf 42}, 
1698 (1979); Phys. Rev. B {\bf 22}, 2099 (1980).

\bibitem{kivelson}
M. I. Salkola and S. A. Kivelson, Phys. Rev. B {\bf 50}, 13962 (1994).

\bibitem{jeckelmann}
E. Jeckelmann, Phys. Rev. B {\bf 57}, 11838 (1998).

\bibitem{fradkin}
E. Fradkin and J. E. Hirsch, Phys. Rev. B {\bf 27}, 1680 (1983);
J. E. Hirsch  and E. Fradkin, Phys. Rev. B {\bf 27}, 4302 (1983).

\bibitem{note4kf}
Since we work on a periodic lattice, $4k_F=2\pi\delta$ and 
$2k_F=\pi(1-\delta)$.

\bibitem{le}
A. Luther and V. J. Emery, Phys. Rev.Lett. {\bf 33}, 589 (1974).

\bibitem{tsuchiizu}
M. Tsuchiizu, A. Furusaki, cond-mat/0109051.

\bibitem{prokofev}
N. V. Prokof'ev, B. V. Svistunov, and I. S. Tupitsyn, Zh. Eks. Teor. Fiz. 
{\bf 64}, 853 (1996); Sov. Phys. JETP {\bf 87}, 310 (1998).

\bibitem{irsse}
A. W. Sandvik, R. R. P. Singh, and D. K. Campbell, Phys. Rev. B 
{\bf 56} 14510 (1997).

\bibitem{kawashima}
N. Kawashima, J. E. Gubernatis, and H. G. Evertz, Phys. Rev. B {\bf 50},
136 (1994). 


\bibitem{pericorr}
A. W. Sandvik and D. J. Scalapino, Phys. Rev. B {\bf 47}, 12333 (1993);
A. W. Sandvik, D. J. Scalapino, and C. Singh, Phys. Rev. B {\bf 48}, 2112 
(1993).


\end{references}
\end{document}